\documentclass[twocolumn, prb, floatfix, superscriptaddress]{revtex4}

\usepackage{graphicx}
\usepackage{amsmath,amssymb}
\usepackage[caption=false]{subfig}
\usepackage{xcolor}

\begin{document}

\title{Condensation in hybrid superconducting cavity-microscopic spins systems with finite-bandwidth drive}  
\author{R. Au-Yeung}
\affiliation{Advanced Technology Institute, University of Surrey, Guildford GU2 7XH, United Kingdom}
\author{M. H. Szyma\'{n}ska}
\affiliation{Department of Physics \& Astronomy, University College London, London WC1E 6BT, United Kingdom}
\author{E. Ginossar}
\thanks{e.ginossar@surrey.ac.uk}
\affiliation{Advanced Technology Institute, University of Surrey, Guildford GU2 7XH, United Kingdom}

\date{\today}

\begin{abstract}
Using Keldysh field theory, we find conditions for non-equilibrium condensation in the open Tavis-Cummings model under a direct finite-bandwidth incoherent cavity drive. Experimentally, we expect the condensation transition to be easily accessible to hybrid superconducting systems coupled to microscopic spins, as well as to many other incoherently driven light-matter systems. In our theoretical analysis, we explicitly incorporate the drive's spectral distribution into the saddle-point description. We show that the injected incoherent photons create a drive-dependent effective coupling between spin-1/2 particles. The condensation transition arises at a critical regime of driving which we can now accurately predict. Our results also provide important guidelines for future quantum simulation experiments of non-equilibrium phases with hybrid devices.
\end{abstract}


\maketitle


\section{Introduction}
 
The last few decades have seen enormous advances in experimental realization and theoretical understanding of quantum condensation in a variety of physical systems. They range from condensed matter models like microcavity exciton-polariton systems \cite{Kasprzak2006,Szymanska2007}, electron-hole plasmas of highly excited semiconductors \cite{Kremp2008}, and quantum magnets \cite{Zapf2014}; to circuit quantum electrodynamics (CQED) devices that demonstrate photon condensation \cite{Marcos2012}, the Dicke states \cite{Wu2017}, and dressed collective qubit states \cite{Fink2009}.

Among these, condensation in light-matter cavity QED systems offers a path into deeper understanding of non-equilibrium phases of matter. Here we show that this phenomenon could be studied with an analogue quantum simulation \cite{Cirac2012,Xiang2013,Georgescu2014,Zagoskin2018}, which offers a highly promising avenue for testing quantitative predictions of non-equilibrium quantum theory \cite{Altman2021}. Analogue simulators can mimic the dynamics of many-body quantum optical systems by reconstructing their Hamiltonian under precisely controlled conditions \cite{Kurizki2015,Fitzpatrick2017,Blais2021}.
In particular, we consider the much studied hybrid quantum system consisting of a superconducting resonator coupled to solid state spin centers \cite{Bonizzoni2018,Mi2018,Clerk2020} as the ideal quantum simulator. Experimental realizations can include a superconducting resonator coupled to NV$^-$ spin centers or other impurities \cite{Schuster2010,Kubo2010,Amsuss2014,Liu2016}. Programmable signal generators and sensitive amplifiers provide highly controllable driving and probing in the microwave regime, even to the extent that quantum state tomography can reconstruct the cavity states of microwave resonators \cite{Kirchmair2013,Weiss2019}. Although CQED setups have well-understood Hamiltonians, the question of how to achieve condensation with these systems has not been addressed before. A key element is energy pumping (driving) which we focus on here. We analyse this system in the presence of an incoherent, finite-bandwidth, typically microwave frequency drive with a general spectral distribution.

The non-equilibrium response to incoherent drive is challenging to analyze because the drive induces interactions between the two-level atoms. This is considered an open problem which was previously approached with an effective phenomenological (or Markovian) description in the system's equations of motion \cite{Sieberer2016} or, where appropriate, with the introduction of fermionic driving \cite{Szymanska2007}. In the latter it was shown that the system can access both condensation and lasing regimes \cite{Kasprzak2006,Deng2010,Byrnes2014,Deveaud2015} which are connected by a smooth crossover \cite{Keeling2005,Keeling2010,Szymanska2013}.

In contrast, here we analyze the condensation transition based on a model with a direct photonic drive. We use Keldysh field theory techniques \cite{Kamenev2011,Sieberer2016} to study the consequences of a bosonic drive and obtain the condensation phase diagram with directly applicable predictions. Technically, our analysis is based on the saddle-point solution of the Keldysh action which is standard procedure for studying condensation phenomena \cite{Sieberer2016,Altland2010}. Note also that because the drive couples to the cavity photon modes, population inversion of the two-level atoms cannot occur. Hence it is physically impossible for the system to behave as a laser.

One of the central aims of our work is to determine the parameter space and photon drive's distribution function where the system can access the condensation regime. We hypothesise that driving the cavity with Markovian noise (flat spectral density) cannot reach the phase transition because it strongly decoheres and thermalizes the system. Indeed, we see that the threshold for condensation is only realistically achievable for a finite bandwidth drive, a case which can be analyzed in our formalism.

In this work, we use the term ``non-Markovian'' driving to mean frequency-dependent driving. We only observe spontaneous condensation when preferentially driving the low frequency part of the spectrum. In the regime of condensation, we observe that the condensation threshold is sensitive to the width of the spectral profile of the drive. Larger widths require higher drive strengths. Hence, extremely large spectral profiles are experimentally unpractical. In fact, it was established in earlier work \cite{Szymanska2001} that condensation cannot be achieved within the Markov approximation, which drives the high and low energy modes with equal strength. The drive would cause strong decoherence that impedes the formation of the condensate \cite{Szymanska2001}. Hence we cannot analyze the problem using a Markovian master equation, which is a more appropriate method for quantum optical problems where energy dependence of decay or drive are irrelevant.


\section{System and model}

In this work, we use Keldysh field theory \cite{Kamenev2011,Sieberer2016} to predict the onset of spontaneous coherence (via non-equilibrium condensation) in many-body light-matter systems under direct incoherent driving with finite bandwidth and amplitude. Standard quantum optics techniques cannot easily treat frequency dependent (non-Markovian) drive. However the field theoretical approach offers a natural language for including general spectral densities and exploring how this affects the steady-state of the system. 

\begin{figure*}[ht]
\centering
\includegraphics[width=\linewidth]{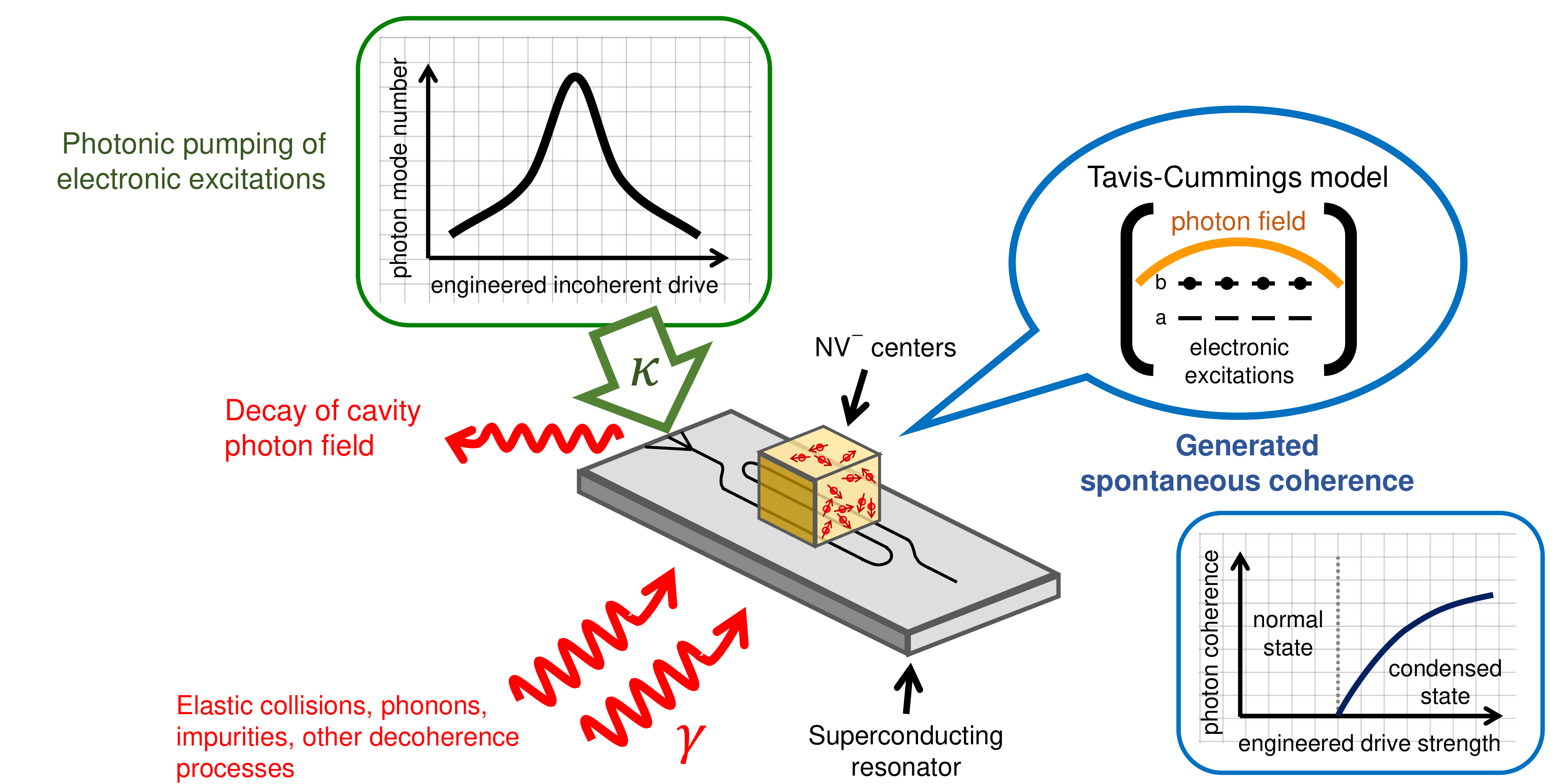}
\caption{Sketch of possible experimental setup uses NV$^-$ vacancy centers (spin-1/2 particles) in diamond coupled to superconducting resonator interacting with different types of environment. Frequency profile of photonic drive characterized by photon mode number $n_B(\omega)$.} 
\label{fig:cavitysystem}
\end{figure*}

We focus on the open Tavis-Cummings model (Fig. \ref{fig:cavitysystem}) featuring an ensemble of spin-1/2 particles (or pseudo-spins representing two-level atoms) which are dipole-coupled to one quantized cavity mode. The cavity is coupled to an incoherent photonic drive. Unlike previous Keldysh-based theoretical studies on microcavity polariton systems \cite{Szymanska2007}, we model the photonic drive as a bosonic reservoir (bath) which feeds incoherent photons into the cavity. We derive an extension to the self-energy of the spins which arise from fluctuations in the cavity field caused by the incoherent drive. In addition, the theory includes a low temperature reservoir to represent the non-radiative environment.

Note that this photonic driving generates a non-equilibrium condensation which differs from the lasing and Dicke-type phase transitions \cite{Kirton2018,Kirton2019}. This work cannot be directly compared with other phenomena such as lasing and super-radiance. It is not an incremental study, but rather a new regime for condensation in the driven-dissipative Tavis-Cummings model which is directly relevant to CQED devices.

The non-equilibrium condensation is characterized by a non-thermal distribution function. The structured spectral density of the drive and its statistical properties are described by a customized distribution function. The signatures of condensation are a sudden appearance of collective phase coherence in the spins and condensation of the cavity photons in a coherent state (Fig. \ref{fig:cavitysystem}). 

We define the Hamiltonian as $\hat{H} = \hat{H}_\textrm{sys} + \hat{H}_\textrm{sys,bath} + \hat{H}_\mathrm{bath}$ ($\hbar=1$) where
\begin{gather}
\hat{H}_\textrm{sys} 
= \sum_{\alpha}
\left[
\epsilon_0(b_\alpha^\dagger b_\alpha-a_\alpha^\dagger a_\alpha) + g_\alpha (\psi b_\alpha^\dagger a_\alpha+\mathrm{H.c.}) 
\right] \nonumber\\
+ \omega_0\psi^\dagger\psi.
\end{gather}
Spin-up and -down states, represented using fermionic annihilation operators $b_\alpha$ and $a_\alpha$ respectively, are dipole-coupled to cavity photon mode $\psi$ with individual spin-photon coupling is $g_\alpha$. We use a fermionic representation of spin-1/2 particles to facilitate a straightforward use of the field theoretical formulation (although the action can be formulated in terms of coherent spin states \cite{Nagaosa1995}, it is technically difficult to treat). For example the operation $b_\alpha^\dagger a_\alpha$ creates a spin-up particle  with energy $\epsilon_0$. The single-occupancy constraint 
\begin{equation}\label{eq:singleoccupancy}
b_\alpha^\dagger b_\alpha + a_\alpha^\dagger a_\alpha = 1
\end{equation}
allows the fermionic representation to properly describe a spin-1/2 particle. This eliminates the unphysical states $\vert \textrm{vacuum} \rangle$ and $b_\alpha^\dagger a_\alpha^\dagger \vert \textrm{vacuum}\rangle$, leaving two possible transitions for the ground $\vert 0 \rangle = a_\alpha^\dagger \vert\textrm{vacuum}\rangle = a_\alpha^\dagger b_\alpha \vert 1 \rangle$ and excited state $\vert 1 \rangle = b_\alpha^\dagger \vert\textrm{vacuum}\rangle = b_\alpha^\dagger a_\alpha \vert 0 \rangle$. The constraint is maintained by considering only states which obey this condition and bath distribution functions as described below. For simplicity, the spin-photon coupling is on resonance i.e. the cavity and atomic frequencies are both $\omega_0$, and there is no detuning in the system i.e. $\omega_0-2\epsilon_0=0$. 
The system is coupled to two baths (reservoirs),
\begin{gather}
\hat{H}_\textrm{sys,bath} 
= \sum_{\alpha,k} 
[ \Gamma_{\alpha,k}(a_\alpha^\dagger A_k + b_\alpha^\dagger B_k + \mathrm{H.c.}) \nonumber\\
+ \zeta_k(\psi^\dagger\Psi_k+\mathrm{H.c.}) ]\label{eq:Hsb} 
\end{gather}
given by $\hat{H}_\mathrm{bath} = \sum_{k} \omega_k^\Gamma ( A_k^\dagger A_k + B_k^\dagger B_k ) + \sum_{k} \omega_k^{\zeta}\Psi_k^\dagger \Psi_k$, where $A_k$ and $B_k$ are fermionic annihilation operators and $\Psi_k$ are bosonic annihilation operators. The fermionic bath modes oscillate at frequencies $\omega_k^\Gamma$ and couple to the spins with coupling constants $\Gamma_{\alpha,k}$. Similarly the bosonic modes have frequencies $\omega_k^\zeta$ and coupling constants $\zeta_k$.

The photonic bath represents a continuum of external electromagnetic modes, for example on a superconducting transmission line in the microwave regime. We assume that the bath modes are populated with photons in a prescribed spectral distribution which can be realized by controllable external sources. (We shall use the terms `photonic' and `bosonic' interchangeably throughout.) We assume that the spectral distribution is broader than the natural linewidth of the cavity mode and possesses random phases for different modes. Under these conditions, this bosonic bath cannot impose any definite phase coherence on the cavity but rather only drive it incoherently. Hence, the condensation transition would be a result of spontaneous $U(1)$ symmetry breaking rather than any externally imposed phase coherence.

It is important to note that under broad (Markovian) spectral distributions we do not expect cavity photons to condense due to the bath's excessive dephasing power\cite{Szymanska2001}. Given these conditions, we expect to find condensation only for cases with a finite bandwidth. Hence we emphasize that the non-Markovian, frequency dependent nature of the bath is crucial.

The fermionic baths, which are simpler to treat in field theory than the bosonic bath, represent the microscopic environment of the spins. These baths are assumed to be cold compared to the spin energy, hence in the absence of any other processes would induce relaxation of the spins to the ground state. The choice of symmetric spectral distributions ensures that on average there are no pair-breaking events \cite{Szymanska2007} i.e. $\langle b_\alpha^\dagger b_\alpha \rangle+ \langle a_\alpha^\dagger a_\alpha \rangle = 1$. Although mathematically we define two fermionic species comprising the fermionic bath, in reality this corresponds to any kind of mechanism that causes spin relaxation and dephasing. 


\section{Methodology} 

Standard procedure of Keldysh field theory in its path-integral formulation \cite{Szymanska2007,Sieberer2016} enables the Keldysh action to be derived. Because the bath fields appear in the action at quadratic level, the functional integral over them can be performed analytically so we simplify the action by removing the bath degrees of freedom using Gaussian integration \cite{Szymanska2007,Dunnett2016}. This yields an effective description in terms of the photon field. 
In the non-equilibrium steady state the two-time Green's functions are time-translational invariant, $\tau=t-t'$. This means we can Fourier transform with respect to $\tau$ into the frequency representation and make some standard assumptions about the bath properties.

Each bath couples equally to the cavity mode so that spontaneous emission is momentum independent. The baths contain many modes, i.e. are much larger than system. They therefore thermalize rapidly compared to any system interactions, and their properties (e.g. distribution function) are unaffected by the system behavior. Coupling strengths $\Gamma_k$ and $\zeta_k$ are smooth functions of momentum $k$. The baths have a dense energy spectrum, so $\sum_k$ summations can be replaced by $\int \mathrm{d}\omega^{\Gamma}$ and $\int \mathrm{d}\omega^{\zeta}$ integrals. We take the Markovian approximation (frequency independence) for the bath density of states ($N^\zeta(\omega^\zeta)=N^\zeta$ and $N^\Gamma(\omega^\Gamma)=N^\Gamma$) and spin-bath coupling ($\zeta(\omega^\zeta)=\zeta$ and $\Gamma(\omega^\Gamma)=\Gamma$).

In summary, this step introduces the cavity-photonic drive coupling strength $\kappa = \pi \zeta^2 N^\zeta$ and cavity-dephasing bath coupling strength $\gamma = \pi \Gamma^2 N^\Gamma$. As there are two different species of fermions that couple to the upper and lower states of the spin-1/2 particle, the latter is considered a relaxation rate that models the microscopic environment. We also introduce three distribution functions: one for the photonic drive $F_\Psi(\omega)$ and two for the dephasing baths \cite{Szymanska2007} $F_{b/a}(\omega)$. Here $F_\Psi(\omega) = 1 + 2n_B(\omega)$ where the photon occupation number $n_B(\omega)$ can have any form. We discuss this later in the calculation.

The dephasing baths' distribution functions are 
\begin{equation}
F_{b/a}(\omega) = 1 - 2 n^{b/a}_F(\omega).
\end{equation}
The occupation functions $n^{b/a}_F(\omega)$ are externally imposed and can be chosen to have any form relevant to a particular physical situation. We choose a thermal reservoir with $n^b_F$ and $n^a_F$ set at equal internal temperature $T_F$ but different chemical potentials $\mu^{b/a}_B$. The system must adhere to the single occupancy constraint (eq. \ref{eq:singleoccupancy}), so the chemical potential of baths $A$ and $B$ are related by $\mu^a_B = -\mu^b_B = -\mu_B$, and the occupations should satisfy \cite{Szymanska2006,Keeling2010} $n^b_F + n^a_F = 1$. Hence we choose
\begin{equation}
F_{b/a}(\omega) = \tanh \dfrac{\beta}{2}(\omega\mp\mu_B)
\end{equation}
with inverse effective temperature $\beta=1/T_F$ and chemical potential $\mu_B$.

In the regime where $\mu_B<0$ ($\mu_B>0$), the lower (upper) state is more likely to be occupied in thermal equilibrium. The $\mu_B\to\mp\infty$ limit corresponds to certainty that the lower/upper state is occupied. For $F_{b/a}$ to behave as a dephasing bath, we choose $\mu_B<0$.

For convenience, we arrange the fermionic fields in a Nambu vector $\phi=[b,a]^T$, with $(1,2)$ indices to represent the fermionic fields under Keldysh rotation \cite{Kamenev2011}. The resulting action is
\widetext\begin{equation}
S = \iint \mathrm{d}t \mathrm{d}t'
\left(
\sum_{\alpha,\alpha'} \bar{\phi}_\alpha(t) \mathbf{G}_\mathbf{0}^{-1}(t,t') \phi_{\alpha'}(t')
+
\bar{\psi}(t) 
\begin{bmatrix}
0 & i\partial_{t'} - \omega_0 - i\kappa
\\
i\partial_{t'} - \omega_0 + i\kappa& 2i\kappa F_\Psi(t-t')
\end{bmatrix}
\psi(t') 
\right)
\end{equation}\twocolumngrid
where $\mathbf{G_0}$ is the Tavis-Cummings Green's function. We use abbreviations $\lambda_{cl,q} = \frac{g_\alpha}{\sqrt{2}} \psi_{cl,q}$ so that we may write 
\widetext\begin{align}
\mathbf{G}_\mathbf{0}^{-1}(t,t')
&= 
\begin{bmatrix}
(i\partial_{t'}\sigma_0 - \epsilon_0\sigma_3 - \lambda_{cl}(t)\sigma_+ - \bar{\lambda}_{cl}(t)\sigma_-) 
\delta_{\alpha,\alpha'}
+ i\gamma\sigma_0 
& 
-(\lambda_{q}(t)\sigma_+ + \bar{\lambda}_{q}(t)\sigma_-)
\delta_{\alpha,\alpha'}
+ 2i\gamma(F_b \sigma_\uparrow + F_a \sigma_\downarrow)
\\
-(\lambda_{q}(t)\sigma_+ + \bar{\lambda}_{q}(t)\sigma_-)
\delta_{\alpha,\alpha'}
&
(i\partial_{t'}\sigma_0 - \epsilon_0\sigma_3 - \lambda_{cl}(t)\sigma_+ - \bar{\lambda}_{cl}(t)\sigma_-)
\delta_{\alpha,\alpha'}
- i\gamma\sigma_0 
\end{bmatrix}.
\end{align}\twocolumngrid

The calculation up to this point has followed the same steps as in previous work \cite{Szymanska2007}. The rest of the calculation is original work.

To introduce the effects of incoherent driving, we separate the photon field into its coherent and incoherent parts by substituting $\psi = \langle\psi\rangle + \delta\psi$ into the action. Gaussian integration removes the fluctuation degrees of freedom $\delta\psi$, and produces a new action containing quartic $\sim\phi^4$ terms \cite{sm}.

Hubbard-Stratonovich decoupling \cite{Altland2010} is required to simplify the quartic terms, at the expense of introducing an auxiliary field $Q^{(c,d)}(t,t')_{\alpha,\alpha'} \sim \langle \phi^c_\alpha(t) \bar{\phi}^d_{\alpha'}(t') \rangle$ which may be expressed in matrix form,
\begin{equation}\label{eq:Q}
\mathbf{Q}
=
\begin{bmatrix}
Q^{(1,1)} & Q^{(1,2)} \\
Q^{(2,1)} & Q^{(2,2)} 
\end{bmatrix}.
\end{equation}
The decoupling makes the action quadratic in $\phi$ and is analogous to the BCS theory of superconductivity. The resulting action indicates that the incoherent drive causes effective spin-spin interactions mediated by photon exchange. This behavior appears in terms of the form \cite{sm}
\widetext\begin{equation}
\sim \iint \mathrm{d}t\mathrm{d}t' \sum_{\alpha,\alpha'}
\bar{\phi}_\alpha(t) G_\psi^{R,A,K}(\pm\tau) Q^{(c,d)}_{\alpha,\alpha'}(\tau) \phi_{\alpha'}(t') \nonumber\\
+ G_\psi^{R,A,K}(\tau) Q^{(c,d)}_{\alpha,\alpha'}(t,t') Q^{(d,c)}_{\alpha',\alpha}(t',t).
\end{equation}\twocolumngrid
$G_\psi$ is the cavity photon Green's function, and $R,A,K$ denote the retarded, advanced and Keldysh components respectively \cite{Kamenev2011,Sieberer2016}. By definition, the retarded and advanced components contain information about the spectral properties whereas the Keldysh component contains statistical properties of photons.

We can show explicitly that the linear terms in $Q$ influence the cavity's Tavis-Cummings dynamics \cite{sm}. The Dyson equation becomes 
\begin{equation}\label{eq:dyson}
\mathbf{G}^{-1} = \mathbf{G}_\mathbf{0}^{-1} - \boldsymbol{\Sigma}, 
\end{equation}
where the Tavis-Cummings part \cite{sm2} is given by
\begin{equation}\label{eq:GJC}
\mathbf{G}_\mathbf{0}^{-1} =
\begin{bmatrix}
(G_0^{-1})^R & (G_0^{-1})^K \\ 0 & (G_0^{-1})^A 
\end{bmatrix}.
\end{equation}
The self-energy 
\begin{equation}\label{eq:sigma}
\boldsymbol{\Sigma}
=
\begin{bmatrix}
\Sigma^{(1,1)} & \Sigma^{(1,2)} \\
\Sigma^{(2,1)} & \Sigma^{(2,2)} 
\end{bmatrix}
\end{equation}
comes directly from the incoherent drive and contains linear terms in $Q$. Note it is written in Keldysh $(1,2)$ and Nambu (particle-hole, or $(b,a)$) space with non-trivial matrix elements $(i,j)=1,2$, where
\widetext\begin{align*}
\Sigma^{(i,i)}_{bb/aa}(\tau) &= -i\dfrac{g^2}{4}
\left[
G_\psi^K(\pm\tau)Q^{(i,i)}_{aa/bb}(\tau)_{\alpha,\alpha'} + 
G_\psi^{R/A}(\pm\tau)Q^{(i,j)}_{aa/bb}(\tau)_{\alpha,\alpha'} + 
G_\psi^{A/R}(\pm\tau)Q^{(j,i)}_{aa/bb}(\tau)_{\alpha,\alpha'}
\right], i \neq j
\\
\Sigma^{(i,j)}_{bb/aa}(\tau) &= -i\dfrac{g^2}{4}
\left[
G_\psi^K(\pm\tau)Q^{(i,j)}_{aa/bb}(\tau)_{\alpha,\alpha'} + 
G_\psi^{R/A}(\pm\tau)Q^{(i,i)}_{aa/bb}(\tau)_{\alpha,\alpha'} + 
G_\psi^{A/R}(\pm\tau)Q^{(j,j)}_{aa/bb}(\tau)_{\alpha,\alpha'}
\right], i \neq j.
\end{align*}\twocolumngrid 
Following standard procedure, we take the narrow bandwidth limit $g_\alpha \to g$ using the same motivations behind similar work\cite{Marchetti2006}, i.e. the individual spin-photon coupling is independent of the spin index $\alpha$. Then we define the collective coupling as $g \to g \sqrt{N}$. For both matrices $\mathbf{G}_\mathbf{0}^{-1}$ (eq. \ref{eq:GJC}) and $\mathbf{\Sigma}$ (eq. \ref{eq:sigma}), $b$ and $a$ define the particle-hole space as
\begin{equation*}
\boldsymbol{\mathcal{D}} = 
\begin{bmatrix}
D_{bb} & D_{ba} \\ D_{ab} & D_{aa}
\end{bmatrix}.
\end{equation*}
The Keldysh Green's function $G^K_\psi$ contains the statistical properties of the cavity photons which, after Fourier transforming into frequency, is
\begin{equation}
G^K_\psi(\omega) = \dfrac{-2i\kappa F_\Psi(\omega)}{(\omega-\omega_0)^2+\kappa^2}
\end{equation}
with system-photonic bath coupling strength $\kappa$. The bath distribution function 
\begin{equation}\label{eq:Fpsi}
F_\Psi(\omega) = 1 + 2n_B(\omega)
\end{equation}
is the energy resolved occupation of (quasi-)particle modes. Its presence in $\mathbf{\Sigma}$ means that the photon environment, with $F_\Psi$ of modes outside the cavity, now affects both the free photon evolution and the development of spontaneous coherence. We will now show that the saddle-point equation includes this distribution function, a significant departure from previous works.

By adding a spin-spin interaction that we derive separately, we can construct an augmented action,
\begin{equation}
S_\textrm{aug} = \iint \mathrm{d}t \mathrm{d}t' \sum_{\alpha,\alpha'} \bar{\phi}_\alpha(t)
\mathbf{G}^{-1}_{\alpha,\alpha'}(t-t')
\phi_{\alpha'}(t') + S_Q + S_\psi
\end{equation}
where $\mathbf{G}$ is the Dyson equation defined in eq. \ref{eq:dyson} and $S_Q$ contains terms quadratic in $Q$.
Then integrating out the fermionic fields $\phi$ produces an effective action written in terms of the photon and auxiliary fields,
\begin{equation}
S_\textrm{aug}^\textrm{eff} = -i \sum_{\alpha} \mathrm{tr} \ln \mathbf{G}^{-1}_{\alpha,\alpha} + S_Q + S_\psi
\end{equation}
where tr traces over all parameters (Keldysh space, particle-hole space, time, $\alpha$ site indices).

It is difficult to invert the $\mathbf{G}^{-1}$ matrix when the photon field $\psi(t)$ has an arbitrary time dependence. Hence we take two further simplifications: the single-frequency ansatz
\begin{equation}
\psi(t) = \psi e^{-i\mu_S t}
\end{equation}
and saddle-point approximation. Since we are interested in non-equilibrium steady states, we take the only time dependence of the photon field to be oscillation at a single frequency $\mu_S$. This leads to explicit time dependence within $\mathbf{G}^{-1}$. We want to search for self-consistent solutions with a steady state, uniform, photon field of the form $\psi(t)$. The $\mu_S$ parameter can also be thought of as the system's chemical potential. The mean-field theory of the non-equilibrium system describes a self-consistent steady state, and inverting the $\mathbf{G}^{-1}$ matrix gives a self-consistent Dyson equation
\begin{equation}
\mathbf{G}(\omega) = \mathbf{G_0}(\omega) + \mathbf{G_0}(\omega) \cdot \boldsymbol{\Sigma}(\omega) \cdot \mathbf{G}(\omega).
\end{equation}

In summary, the effects of such a time dependence appear in two places: in the time derivative terms, which lead to the energy shifts $\omega_0 \to \omega_0-\mu_S$ and $\epsilon_0 \to \epsilon_0-\mu_S/2$, and in a gauge transformation of the bath functions $F_\Psi(\omega) \to F_\Psi(\omega+\mu_S)$ and $F_{b/a}(\omega) \to F_{b/a}(\omega\pm\mu_S/2)$. 

It is customary to assume that the dominant contribution to the quantum partition function arises from the configurations $\psi$ and $Q$ which minimise the total action. This requires taking the functional derivative of the action with respect to $\bar{\psi}_q$, then setting $\psi_q=0$:
\begin{equation}
\dfrac{\delta S_\textrm{aug}^\textrm{eff}}{\delta \bar{\psi}_q} \bigg\vert_{\psi_q=0} = 0.
\end{equation}
We get the saddle point equation
\begin{equation}\label{eq:spe}
(\omega_0-\mu_S-i\kappa) \langle\psi\rangle 
= i\dfrac{g}{2} \sum_\alpha \mathrm{tr} \; G^K_{ba}(\tau)_\alpha,
\end{equation}
where tr traces over all parameters. Note that there always exists a solution to the saddle point equations where the quantum part is zero, which corresponds to the purely classical limit. Hence when $\bar{\psi}_q=0$, then $\delta S_\textrm{aug}^\textrm{eff}/\delta \bar{\psi}_{cl} = 0$ trivially. 

Similarly, varying the action with respect to $Q$,
\begin{equation}
\dfrac{\delta S_\textrm{aug}^\textrm{eff}}{\delta Q^{(c,d)}} = 0
\end{equation}
gives
\begin{equation}\label{eq:Qspe}
\mathbf{Q}_{\alpha,\alpha'}(\tau) = -2 \mathbf{G}_{\alpha,\alpha'}(\tau).
\end{equation}
Hence we can redefine the auxiliary field with respect to the Dyson equation. Substituting $Q$ into the Dyson equation (eq. \ref{eq:dyson}) and
following causality arguments \cite{Kamenev2011} inherent to the formalism allows us to simplify the self-energy. Hence we substitute $Q^{(1,1)} = Q^R$, $Q^{(1,2)} = Q^K$, $Q^{(2,1)} = 0$, and $Q^{(2,2)} = Q^A$. We rewrite the self-energy in matrix form,
\begin{equation*}
\mathbf{\Sigma} = 
\begin{bmatrix}
\Sigma^R & \Sigma^K \\ 0 & \Sigma^A
\end{bmatrix}
\end{equation*}
where $\Sigma^{(1,1)}_{bb} = \Sigma^R_{bb}$, $\Sigma^{(1,2)}_{bb} = \Sigma^K_{bb}$, $\Sigma^{(2,1)}_{bb} = \Sigma^{(2,1)}_{aa} = 0$, $\Sigma^{(2,2)}_{bb} = \Sigma^A_{bb}$, $\Sigma^{(1,1)}_{aa} = \Sigma^R_{aa}$, $\Sigma^{(1,2)}_{aa} = \Sigma^K_{aa}$, $\Sigma^{(2,2)}_{aa} = \Sigma^A_{aa}$.

The mean-field properties of a driven cavity are described by a complex analogue of the Gross-Pitaevskii equation in the BEC regime or equivalently the gap equation in the BCS regime \cite{Keeling2005}. The real part of the self-consistent saddle-point equation (eq. \ref{eq:spe}) relates the coherent field to the system's nonlinear susceptibility, as in the case of equilibrium condensation, while the imaginary part reflects how the gain and decay are balanced, as in a laser. The equation contains two unknown parameters that characterize the system: the coherent photon field $\langle\psi\rangle$ and its common oscillation frequency $\mu_S$. The non-equilibrium Green's function is defined as $G^K_{ba}(t-t')_\alpha = -i \langle a_{cl}^\dagger(t)_\alpha b_{cl}(t')_\alpha \rangle$, where $a_{cl} = (a_f+a_b)/\sqrt{2}$ and $f$ and $b$ are the forward and backward branches of the Keldysh time contour \cite{Szymanska2007,Kamenev2011}. 

As the name implies, the mean-field description (eq. \ref{eq:spe}) should only contain details of the coherent photons, i.e. the saddle-point approximation filters out the average field and discards any fluctuations. In contrast, our strategy is to integrate over fluctuations $\delta\psi$ so that the effects of incoherent photons appear in the action via corrections to the self-energy. Then after taking the saddle-point approximation, the photonic drive distribution function contributes at the mean field level.

The saddle-point equation itself is complex and hence contains two nonlinear simultaneous equations (real and imaginary parts). It is necessary to use numerical methods to solve for two unknowns $\mu_S$ and $\langle\psi\rangle$. We use numerical solvers based on trust region techniques \cite{sm,Nocedal2006}, a simple yet robust method in optimization problems. For a multi-valued function $Y$, the goal is to search for some vector $x$ that satisfies the $Y(x)=0$ constraint or $Y(x)\approx0$ if a defined boundary condition is satisfied. It is necessary to locally approximate the least-squares objective and constraint functions with some simple functions that are easy to optimize. The trust region algorithm uses a quadratic approximation of the objective function, then optimizes the functions within some localized trust region around a point $x'$ to obtain a candidate step $y$. If $y$ is an improvement on the previous iteration of $y$, the algorithms moves closer to the centre of the trust region. If not, the approximations and trust region is adjusted.


\section{Non-equilibrium condensation}

By considering the external photon mode distribution function (eq. \ref{eq:Fpsi}) with specific frequency profiles, we now show that varying the drive parameters does indeed permit the development of spontaneous coherence. For this work, we set the photon occupation number $n_B$ to be a Lorentzian
\begin{equation}
n_B(\omega) = h \dfrac{2\Omega}{[\omega-(\xi-\mu_S)]^2+\Omega^2}, 
\end{equation}
centered at $\xi$ (resonance), with a width $\Omega$ (drive profile width) and amplitude $h$. For simplicity, all parameters are scaled to $g=1$. As discussed above, we assume a cold fermionic (dephasing) bath and hence we fix its internal effective temperature to be $T_F=0.1g$. We also let $\epsilon_0=0$ with zero detuning $\omega_0-2\epsilon_0=0$.

The following phase diagrams show the system properties such as the common oscillation frequency of cavity photons $\mu_S$
, photon coherence $\langle\psi\rangle$, absolute value of polarization $\vert\langle a^\dagger b \rangle\vert = \sqrt{(\omega_0-\mu_S)^2+\kappa^2} \vert\psi\vert$ (fraction of condensed spins), and  density of spin excitations $\rho=\frac{1}{2}\sum_\alpha (1 + b_\alpha^\dagger b_\alpha - a_\alpha^\dagger a_\alpha)$. The minimum $\rho=0$ corresponds to all spins in the ground state, whereas $\rho=1$  corresponds to all spins in the excited state, i.e. maximum inversion. In between, there is lasing when $\rho>0.5$, non-equilibrium condensation when $\rho<0.5$, and a smooth crossover connecting the two regimes. 

\begin{figure}[ht]
\centering
\includegraphics[width=1\linewidth]{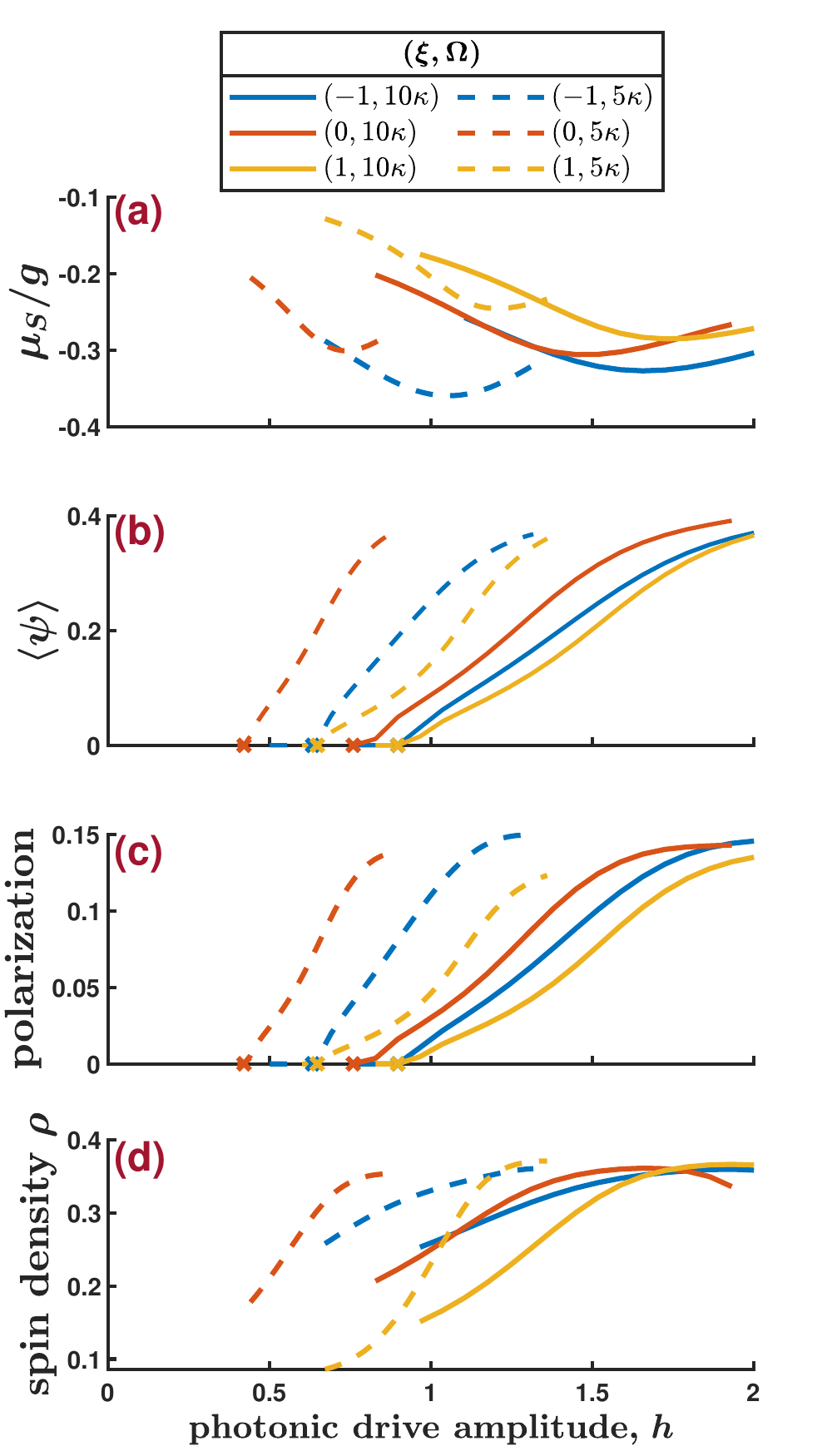}
\caption{System observables of driven system. Parameters $[\mu_B, \kappa, \gamma] = [-0.5, 0.25, 0.75]g$ and Lorentzian distribution $n_B$ with  $[\xi,\Omega]$. Solid lines indicate drive profile width $\Omega=10\kappa$, dashed lines $\Omega=5\kappa$. Larger $\Omega$ requires stronger driving (larger $h$) for system to condense. Driving off-resonance $\xi\neq0$ also requires larger $h$ to reach condensation transition. Markers 'x' on $\langle\psi\rangle$ and polarization graphs indicate where the condensation transition occurs.}
\label{fig:phasediagram1}
\end{figure}

\begin{figure}[ht]
\centering
\includegraphics[width=1\linewidth]{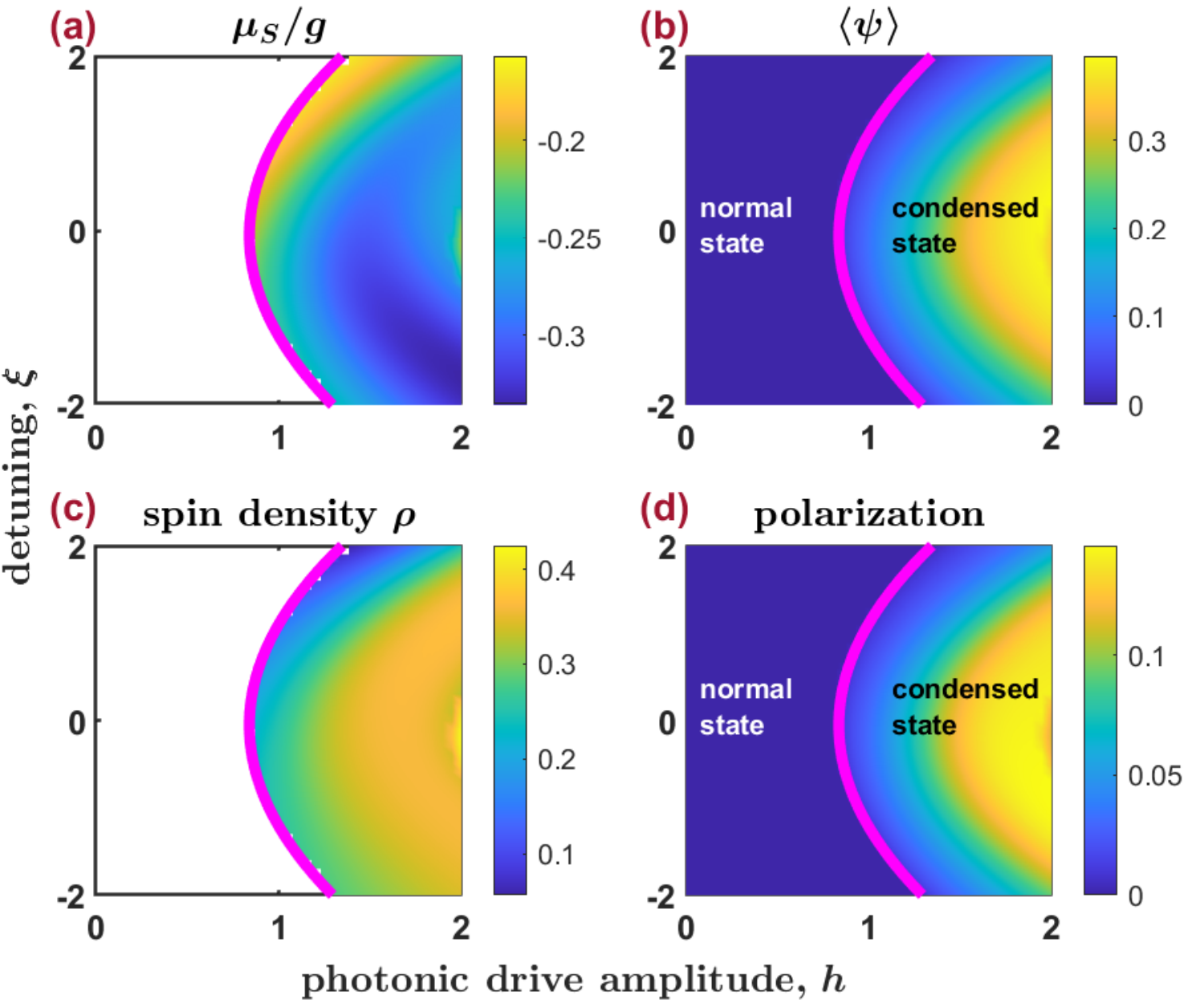}
\caption{Driven system with Lorentzian distribution $n_B$ of external photons, parameters $[\mu_B,\kappa,\gamma,\Omega] = [-0.5,0.25,0.75,10\kappa]g$. Pink lines indicate the phase boundary where condensation transition occurs.} \label{fig:phasediagram3}
\end{figure}

\begin{figure}[ht]
\centering
\includegraphics[width=1\linewidth]{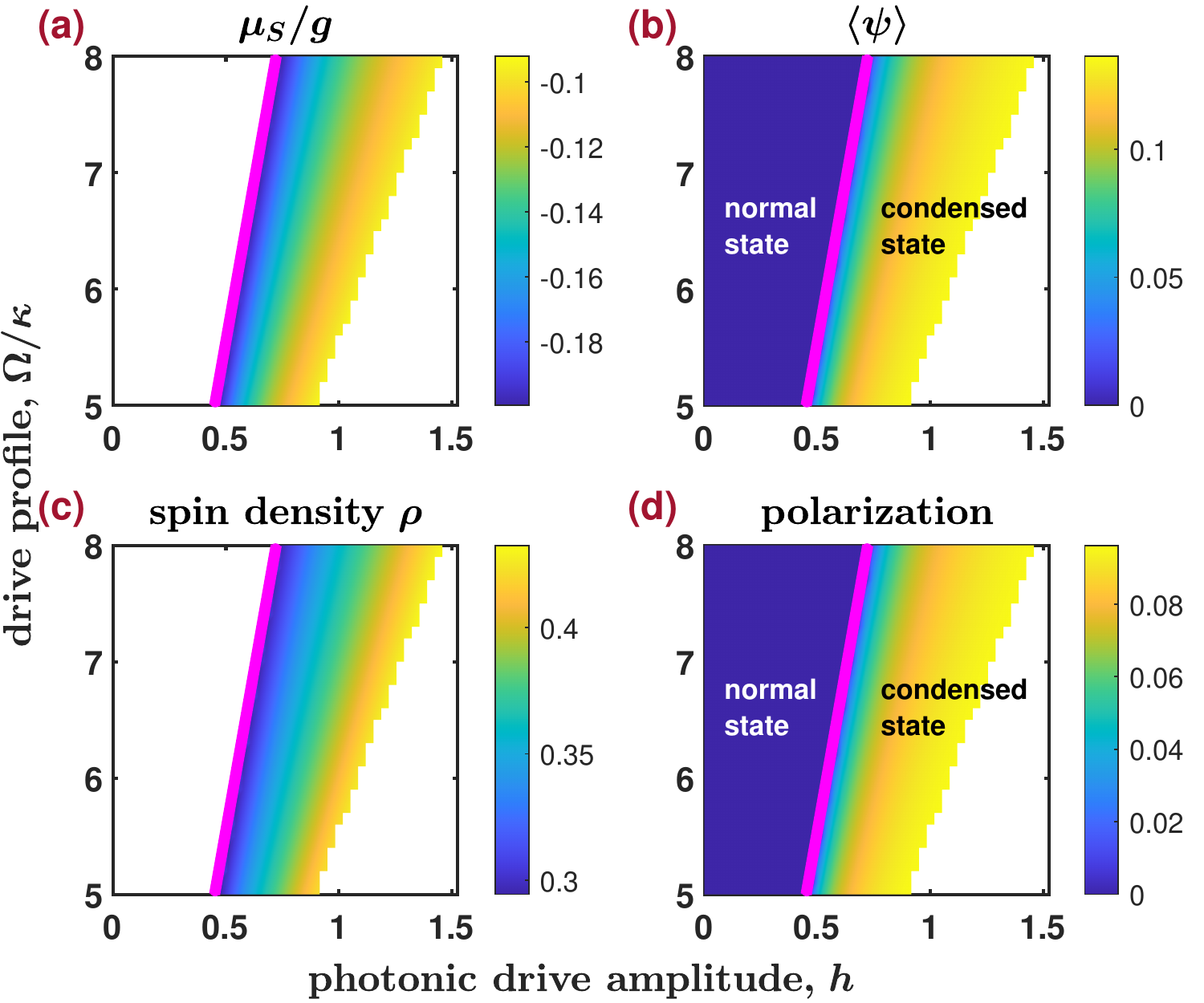}
\caption{Driven system with Lorentzian distribution $n_B$ of external photons, parameters $[\mu_B,\kappa,\gamma,\xi] = [-2,0.7,0.3,0]g$. Pink lines indicate phase boundary of condensation transition.} 
\label{fig:phasediagram2}
\end{figure}

Fig. \ref{fig:phasediagram1}, \ref{fig:phasediagram3} and \ref{fig:phasediagram2} show the system's steady state observables upon increasing the photonic drive strength $h$. The condensation transition occurs when polarization and $\langle\psi\rangle$ become finite at some critical drive strength. In particular, setting chemical potential $\mu_B/g<0$ ensures that the detuning bath causes spontaneous decay processes instead of populating the higher energy spin state.
As expected, a larger driving amplitude $h$ is required to reach the condensation transition when the drive is off resonance. The transition occurs at the same critical drive strength when $\xi = \pm 1$, indicating a symmetry in the system. However the different values for $\mu_S/g$ and $\langle\psi\rangle$ suggest that off-resonance driving affects the condensed state differently when $\xi = \pm1$ (Fig. \ref{fig:phasediagram1}). Decreasing the drive profile width $\Omega$ increases the number of photons per mode around the $\xi$ frequency. This effectively allows the condensation transition to occur at a smaller drive amplitude. We also see that larger $\Omega$ increases the condensation threshold. Hence we do not expect the system to condense in the Markovian limit (i.e. large $h$ and $\Omega\to\infty$).
At some upper limit of the drive amplitude, the numerical method fails. This possibly demonstrates that there is a region of inaccessible parameters. It may be caused by a numerical issue, or the system could be entering some unknown state that displays exotic, perhaps chaotic dynamics. Since the drive amplitude controls the level of noise, it is also possible that the dephasing overpowers the spin-photon dipole coupling and interrupts the condensation processes. 

This inaccessible region is more obvious in a two-dimensional phase diagram. Fig. \ref{fig:phasediagram3} uses the same parameters as in Fig. \ref{fig:phasediagram1}. Here we sweep the drive amplitude $\xi$ and Lorentzian detuning. The phase diagram has an almost symmetric phase boundary along the $\xi=0$ axis. It is clear that as the drive detuning increases and becomes more off resonance, the drive amplitude must increase in order for the system to condense. When the system is in a condensed state, a large detuning and strong driving ($h=2$) regime results in non-trivial differences in the system observables when $\xi=\pm2$. Similarly, Fig. \ref{fig:phasediagram2} shows that as drive profile width $\Omega$ increases (which decreases the number of photons per mode around $\xi=0$), the system remains in a condensed state for larger ranges of the drive amplitude before entering the region of inaccessible parameters. 
Note also that the spin density is always $\rho<0.5$. In this regime the system is either in the condensed or normal state. The density never reaches $\rho \geq 0.5$ which shows that the system cannot behave as a laser. Instead as drive amplitude $h$ increases, the rate of increase in photon coherence $\langle\psi\rangle$, polarization, and density start levelling off before the system enters the region of inaccessible parameters.
In addition to taking a Lorentzian frequency distribution, we examined a frequency independent or Markovian drive. We used parameters that would be physically sensible for a system consisting of superconducting resonator coupled to solid state spin-1/2 particles. We explored different regimes involving combinations of $\gamma\leq\kappa$ or $\gamma>\kappa$, and $\mu_B\leq0$ or $\mu_B>0$, and gradually increased the amount of Markovian noise (drive amplitude). We found no reliable indications of condensation. This is an expected result, as the above phase diagrams show how increasing the bandwidth $\Omega$ requires increasing the driving strength for the system to reach the condensation transition.
In Szyma\'{n}ska's work using Maxwell-Bloch equations \cite{Szymanska2001}, when all decay baths are Markovian and finite, the system transitions from condensate-like (no inversion) at zero decay to one requiring inversion. With frequency independent driving and decay, there must be inversion for phase transition to lasing. One major difference is that Szyma\'{n}ska's work involved driving the spins directly by coupling to the bath modes, so strong driving could easily produce inversion. However if instead we drive the incoherent photons with large energies up to infinite energies, it is highly unlikely to induce population inversion or condensation.

\section{Summary and outlook}

We used Keldysh path-integral field theory to study the open Tavis-Cummings model. This allowed us to demonstrate the possibility of condensation and spontaneous coherence under conditions of finite-bandwidth incoherent driving with a Lorentzian frequency profile. It is crucial to use a frequency dependent drive biased towards the lower frequencies. By exclusively driving the low energy modes, we create ideal conditions for developing spontaneous coherence. The resulting condensation is caused by the photonic drive which induces an effective spin-spin coupling mediated by photons. This phenomenon appears explicitly in the field theory and saddle-point equation. The interactions arising from this incoherent drive may cause non-linear effects that are not easily seen in other systems. Because the phase diagrams contain areas where there are no solutions, the obvious next step would be to investigate the possibility of different phases in those regimes, and to explore the influence of fluctuations beyond the mean-field physics reported here\cite{sm3}. 

This theory can be used directly in state-of-the-art CQED experiments. Given the microscopic system parameters and drive profile, quantitative predictions can be made now for finding the non-equilibrium condensation transition, and exploring novel non-equilibrium phases with finite-bandwidth drives. Since it is challenging to find a parameter space where we can observe condensation, further exploration would be extremely useful, as would understanding the behavior seen in the phase diagrams. By applying our findings to analogue quantum simulators, this could pave the way for fundamental research into many-body physics and powerful new technological applications.
%


\acknowledgments

R. A. acknowledges funding from University of Surrey's University Research Scholarship and Overseas Research Scholarship, and Universities New Zealand's Edward \& Isabel Kidson Scholarship. E. G. acknowledges support from the European Commission under project HiTIMe and EPSRC grants (Grant No. EP/I026231/1 and EP/L02263X/1). We also acknowledge useful discussions with J. M. J. Keeling. The data underlying this work are available without restriction. Details of the data and how to request access are available from the University of Surrey publications repository.


\vbox{\vskip0cm} \appendix\setcounter{figure}{0} \renewcommand{\thefigure}{B%
\arabic{figure}}

\section{Keldysh field theory calculation}\label{app:keldysh}

This Appendix presents technical details related to the analytical procedure. The bulk of the analysis involves Keldysh field theory in its functional integral formulation. We use the Hubbard-Stratonovich transformation to simplify the action, at the expense of introducing an auxiliary field. 

The simplification ultimately allows the photon distribution to enter the mean-field description. In the Keldysh formalism, this is evident in the extension to the self-energy in the Dyson equation. Further simplification involves assuming that the cavity photons oscillate at a single frequency (taking the single-frequency ansatz). The final analytical step is mean field analysis. This requires taking the functional derivative with respect to the photon field to produce a complex Gross-Pitaevskii equation, and the auxiliary field which makes it possible to write it in terms of the Dyson equation.

\subsection{Building a Tavis-Cummings system}

After constructing the Keldysh action using standard procedures \cite{Szymanska2007}, we arrange the fermionic fields into a Nambu vector $\phi = [b,a]^T$ and $\phi = [\bar{b},\bar{a}]$. Then the quantum partition function is
\begin{equation}
Z = \mathcal{N} \int \sum_{\alpha,k} \mathbf{D}[\bar{\phi}_\alpha,\phi_\alpha,\bar{\psi},\psi,\bar{A}_k,A_k,\bar{B}_k,B_k,\bar{\Psi}_k,\Psi_k] \mathrm{e}^{iS}
\end{equation}
with normalisation constant $\mathcal{N}$, Keldysh action \begin{equation}
S = \int_C \mathrm{d}t \, \langle \Lambda(t) \vert i\partial_t - \hat{H} \vert \Lambda(t) \rangle,
\end{equation}
and coherent states specified by the fields in
\widetext\begin{equation}
\vert \Lambda(t) \rangle = \vert \bar{\phi}_\alpha(t), \phi_\alpha(t), \bar{\psi}(t), \psi(t), \bar{A}_k(t), A_k(t), \bar{B}_k(t), B_k(t), \bar{\Psi}_k(t), \Psi_k(t) \rangle
\end{equation}\twocolumngrid
on Schwinger-Keldysh contour $C$. In the Keldysh formalism, the system evolves from the distant past ($t=-\infty$) to the distant future ($t=\infty$) on the forwards branch of a closed time contour and returns along the backwards branch. For bosonic fields, the time evolution on the two branches is written in terms of separate fields on each branch $\psi_{f,b}$ and rotated into the classical/quantum basis via the Keldysh rotation \cite{Kamenev2011,Sieberer2016},
\begin{gather}
\psi_{cl/q} = \dfrac{1}{\sqrt{2}}(\psi_f\pm\psi_b), \\
\bar{\psi}_{cl/q} = \dfrac{1}{\sqrt{2}}(\bar{\psi}_f\pm\bar{\psi}_b).
\end{gather} 
Similarly the fermionic fields $\phi$ are rotated into the Larkin-Ovchinnikov basis \cite{Kamenev2011}
\begin{gather}
\phi^{(1,2)} = \dfrac{1}{\sqrt{2}}(\phi_f\pm\phi_b), \\
\bar{\phi}^{(1,2)} = \dfrac{1}{\sqrt{2}}(\bar{\phi}_f\mp\bar{\phi}_b).
\end{gather}

Integrating out the baths generates a Keldysh action in terms of the photon and fermion fields, $S = S_\phi + S_\psi$ with
\begin{equation}\label{eq:Sphi}
S_\phi = \iint \mathrm{d}t \mathrm{d}t' \sum_{\alpha,\alpha'} \bar{\phi}_\alpha(t) \mathbf{G}_\mathbf{0}^{-1}(t,t') \phi_{\alpha'}(t'),
\end{equation}
and Tavis-Cummings Green's function
$\mathbf{G}_\mathbf{0}^{-1}$.

The photon action is
\begin{equation}
S_\psi = \iint \mathrm{d}t \mathrm{d}t' \, \bar{\psi}(t) \mathbf{G}_{\boldsymbol{\psi}}^{-1}(t,t') \psi(t') 
\end{equation}
with photon Green's function
\begin{align}
\mathbf{G}_{\boldsymbol{\psi}}^{-1}(t,t') &=
\begin{bmatrix}
0 & (G_\psi^{-1})^A \\ (G_\psi^{-1})^R & (G_\psi^{-1})^K 
\end{bmatrix} 
\\
&=
\begin{bmatrix}
0 & i\partial_{t'} - \omega_0 - i\kappa
\\
i\partial_{t'} - \omega_0 + i\kappa& 2i\kappa F_\Psi(t-t')
\end{bmatrix}.
\end{align}
As an aside: The Fourier transform \cite{Mahan1990} is defined for some arbitrary complex function $x(\tau)$ in time $\tau=t-t'$,
\begin{gather}
X(\omega) = \int_{-\infty}^{\infty} \mathrm{d}\tau \; \mathrm{e}^{i\omega\tau} x(\tau), \\ 
x(\tau) = \int_{-\infty}^{\infty} \dfrac{\mathrm{d}\omega}{2\pi} \mathrm{e}^{-i\omega\tau} X(\omega).
\end{gather} 
The single frequency ansatz makes it possible to invert $\mathbf{G}_\mathbf{0}^{-1}$. We let $\psi(t) = \psi \mathrm{e}^{-i\mu_St}$, hence the photon field $\psi$ time dependence\cite{Szymanska2007} occurs as oscillation at a single frequency $\mu_S$. Then using the Fourier transform gives the Tavis-Cummings terms in frequency domain,

\begin{equation}
\mathbf{G_0} =
\begin{bmatrix}
(G^R_{bb})_0 & (G^R_{ba})_0 & (G^K_{bb})_0 & (G^K_{ba})_0 \\
(G^R_{ab})_0 & (G^R_{aa})_0 & (G^K_{ab})_0 & (G^K_{aa})_0 \\
0 & 0 & (G^A_{bb})_0 & (G^A_{ba})_0 \\
0 & 0 & (G^A_{ab})_0 & (G^A_{aa})_0 
\end{bmatrix}
\end{equation}
where 
\widetext\begin{gather}
G^{R/A}_0(\omega) = 
\dfrac{(\omega \pm i\gamma)\sigma_0 + \epsilon\sigma_3 + g_\alpha(\psi\sigma_+ + \bar{\psi}\sigma_-)}{\omega^2 - E_\alpha^2 \pm 2i\gamma\omega - \gamma^2} 
\\
G^K_{bb/aa}(\omega)_0 = 
-2i\gamma
\dfrac{((\omega\pm\epsilon)^2+\gamma^2)F_{b/a}(\omega) + g_\alpha^2\vert\psi_f\vert^2 F_{a/b}(\omega)}{((\omega-E_\alpha)^2+\gamma^2)((\omega+E_\alpha)^2+\gamma^2)} 
\\
G^K_{ba}(\omega)_0 = -G^K_{ab}(\omega)_0^* =
-2i\gamma g_\alpha \psi_f
\dfrac{(\omega+\epsilon+i\gamma)F_b(\omega) + (\omega-\epsilon-i\gamma) F_a(\omega)}{((\omega-E_\alpha)^2+\gamma^2)((\omega+E_\alpha)^2+\gamma^2)}, 
\end{gather}\twocolumngrid
where $\epsilon = \epsilon_0-\mu_S/2$ and $E_\alpha=\sqrt{\epsilon^2+g_\alpha^2\vert\psi_f\vert^2}$. (When taking the saddle-point approximation of the action $S=S_\phi+S_\psi$, this requires setting $\psi_q=0$ so that $\psi_f=(\psi_{cl}+\psi_q)/\sqrt{2}=\psi_{cl}/\sqrt{2}$, and we recover the $RAK$ Keldysh matrix structure.)

Similarly the photon Green's functions are
\begin{gather}
\mathbf{G}_{\boldsymbol{\psi}} =
\begin{bmatrix}
G_\psi^K & G_\psi^R \\ G_\psi^A & 0
\end{bmatrix}
\\
G^{R/A}_\psi(\omega) = \dfrac{1}{\omega-\omega_0 \pm i\kappa}, \\
G^{K}_\psi(\omega) = \dfrac{-2i\kappa F_\Psi(\omega)}{(\omega-\omega_0)^2 +\kappa^2}. 
\end{gather}

\subsection{Addition of incoherent photon field}

After integrating out the bath degrees of freedom, we separate the cavity photon field into its coherent and incoherent parts $\psi = \langle \psi \rangle + \delta\psi$, where the latter component is the result of an externally imposed optical drive.

Again using Gaussian integration to remove the fluctuation fields produces several terms. The analysis shows that the optical driving produces effective spin-spin coupling mediated by exchange of photons,
\widetext\begin{align}
S_{\phi\phi} 
&= 
-\dfrac{g^2}{2} \iint \mathrm{d}t \mathrm{d}t' \sum_{\alpha,\alpha'} 
\biggr( 
\left( 
\bar{b}_\alpha^{(1)} a_\alpha^{(1)} + \bar{b}_\alpha^{(2)} a_\alpha^{(2)}
\right)_t 
G_\psi^K(t,t') 
\left( 
\bar{a}_{\alpha'}^{(1)} b_{\alpha'}^{(1)} + \bar{a}_{\alpha'}^{(2)} b_{\alpha'}^{(2)}
\right)_{t'} \nonumber\\
&\hspace{8em} 
+ \left( 
\bar{b}_\alpha^{(1)} a_\alpha^{(1)} + \bar{b}_\alpha^{(2)} a_\alpha^{(2)}
\right)_t 
G_\psi^R(t,t') 
\left( 
\bar{a}_{\alpha'}^{(2)} b_{\alpha'}^{(1)} + \bar{a}_{\alpha'}^{(1)} b_{\alpha'}^{(2)}
\right)_{t'} \nonumber\\
&\hspace{8em} 
+ \left( 
\bar{b}_\alpha^{(1)} a_\alpha^{(2)} + \bar{b}_\alpha^{(2)} a_\alpha^{(1)}
\right)_t 
G_\psi^A(t,t') 
\left( 
\bar{a}_{\alpha'}^{(1)} b_{\alpha'}^{(1)} + \bar{a}_{\alpha'}^{(2)} b_{\alpha'}^{(2)}
\right)_{t'}
\biggr).
\end{align}\twocolumngrid
After reorganising and symmetrising the action, we use the Hubbard-Stratonovich transformation \cite{Altland2010} to decouple the quartic $\sim\phi^4$ terms. The process introduces an additional auxiliary field $Q^{cd}(t,t')_{\alpha,\alpha'} \sim \langle \phi^c_\alpha(t) \bar{\phi}^d_{\alpha'}(t') \rangle$:
\widetext{\allowdisplaybreaks\begin{align}
&S_{\phi\phi} 
= 
\iint \mathrm{d}t \mathrm{d}t' \sum_{\alpha,\alpha'} 
\dfrac{g^2}{8} G_\psi^K(t,t') Q^{(1,1)}_{aa}(t,t')_{\alpha,\alpha'} Q^{(1,1)}_{bb}(t',t)_{\alpha',\alpha}
\nonumber \\
&\hspace{6em} 
- i \dfrac{g^2}{4} 
\left(
\bar{b}_\alpha^{(1)}(t) G_\psi^K(t,t') Q^{(1,1)}_{aa}(t,t')_{\alpha,\alpha'} b_{\alpha'}^{(1)}(t')
+
\bar{a}_\alpha^{(1)}(t) G_\psi^K(t',t) Q^{(1,1)}_{bb}(t,t')_{\alpha,\alpha'} a_{\alpha'}^{(1)}(t')
\right) \nonumber \\
&\hspace{4em} 
+
\dfrac{g^2}{8} G_\psi^K(t,t') Q^{(1,2)}_{aa}(t,t')_{\alpha,\alpha'} Q^{(2,1)}_{bb}(t',t)_{\alpha',\alpha}
\nonumber \\
&\hspace{6em} 
- i \dfrac{g^2}{4} 
\left(
\bar{b}_\alpha^{(1)}(t) G_\psi^K(t,t') Q^{(1,2)}_{aa}(t,t')_{\alpha,\alpha'} b_{\alpha'}^{(2)}(t')
+
\bar{a}_\alpha^{(2)}(t) G_\psi^K(t',t) Q^{(2,1)}_{bb}(t,t')_{\alpha,\alpha'} a_{\alpha'}^{(1)}(t')
\right) \nonumber \\
&\hspace{4em} 
+ 
\dfrac{g^2}{8} G_\psi^K(t,t') Q^{(2,1)}_{aa}(t,t')_{\alpha,\alpha'} Q^{(1,2)}_{bb}(t',t)_{\alpha',\alpha}
\nonumber \\
&\hspace{6em} 
- i \dfrac{g^2}{4} 
\left(
\bar{b}_\alpha^{(2)}(t) G_\psi^K(t,t') Q^{(2,1)}_{aa}(t,t')_{\alpha,\alpha'} b_{\alpha'}^{(1)}(t')
+
\bar{a}_\alpha^{(1)}(t) G_\psi^K(t',t) Q^{(1,2)}_{bb}(t,t')_{\alpha,\alpha'} a_{\alpha'}^{(2)}(t')
\right)
\nonumber \\
&\hspace{4em} 
+ 
\dfrac{g^2}{8} G_\psi^K(t,t') Q^{(2,2)}_{aa}(t,t')_{\alpha,\alpha'} Q^{(2,2)}_{bb}(t',t)_{\alpha',\alpha}
\nonumber \\
&\hspace{6em} 
- i \dfrac{g^2}{4} 
\left(
\bar{b}_\alpha^{(2)}(t) G_\psi^K(t,t') Q^{(2,2)}_{aa}(t,t')_{\alpha,\alpha'} b_{\alpha'}^{(2)}(t')
+
\bar{a}_\alpha^{(2)}(t) G_\psi^K(t',t) Q^{(2,2)}_{bb}(t,t')_{\alpha,\alpha'} a_{\alpha'}^{(2)}(t')
\right)
\nonumber \\
&\hspace{4em} 
+ 
\dfrac{g^2}{8} G_\psi^R(t,t') Q^{(1,2)}_{aa}(t,t')_{\alpha,\alpha'} Q^{(1,1)}_{bb}(t',t)_{\alpha',\alpha}
\nonumber \\
&\hspace{6em} 
- i \dfrac{g^2}{4} 
\left(
\bar{b}_\alpha^{(1)}(t) G_\psi^R(t,t') Q^{(1,2)}_{aa}(t,t')_{\alpha,\alpha'} b_{\alpha'}^{(1)}(t')
+
\bar{a}_\alpha^{(2)}(t) G_\psi^R(t',t) Q^{(1,1)}_{bb}(t,t')_{\alpha,\alpha'} a_{\alpha'}^{(1)}(t')
\right)
\nonumber \\
&\hspace{4em} 
+ 
\dfrac{g^2}{8} G_\psi^R(t,t') Q^{(1,1)}_{aa}(t,t')_{\alpha,\alpha'} Q^{(2,1)}_{bb}(t',t)_{\alpha',\alpha}
\nonumber \\
&\hspace{6em} 
- i \dfrac{g^2}{4} 
\left(
\bar{b}_\alpha^{(1)}(t) G_\psi^R(t,t') Q^{(1,1)}_{aa}(t,t')_{\alpha,\alpha'} b_{\alpha'}^{(2)}(t')
+
\bar{a}_\alpha^{(1)}(t) G_\psi^R(t',t) Q^{(2,1)}_{bb}(t,t')_{\alpha,\alpha'} a_{\alpha'}^{(1)}(t')
\right)
\nonumber \\
&\hspace{4em} 
+
\dfrac{g^2}{8} G_\psi^R(t,t') Q^{(2,2)}_{aa}(t,t')_{\alpha,\alpha'} Q^{(1,2)}_{bb}(t',t)_{\alpha',\alpha}
\nonumber \\
&\hspace{6em} 
- i \dfrac{g^2}{4} 
\left(
\bar{b}_\alpha^{(2)}(t) G_\psi^R(t,t') Q^{(2,2)}_{aa}(t,t')_{\alpha,\alpha'} b_{\alpha'}^{(1)}(t')
+
\bar{a}_\alpha^{(2)}(t) G_\psi^R(t',t) Q^{(1,2)}_{bb}(t,t')_{\alpha,\alpha'} a_{\alpha'}^{(2)}(t')
\right)
\nonumber \\
&\hspace{4em} 
+
\dfrac{g^2}{8} G_\psi^R(t,t') Q^{(2,1)}_{aa}(t,t')_{\alpha,\alpha'} Q^{(2,2)}_{bb}(t',t)_{\alpha',\alpha}
\nonumber \\
&\hspace{6em} 
- i \dfrac{g^2}{4} 
\left(
\bar{b}_\alpha^{(2)}(t) G_\psi^R(t,t') Q^{(2,1)}_{aa}(t,t')_{\alpha,\alpha'} b_{\alpha'}^{(2)}(t') 
+
\bar{a}_\alpha^{(1)}(t) G_\psi^R(t',t) Q^{(2,2)}_{bb}(t,t')_{\alpha,\alpha'} a_{\alpha'}^{(2)}(t')
\right) \nonumber \\
&\hspace{4em} 
+ 
\dfrac{g^2}{8} G_\psi^A(t,t') Q^{(2,1)}_{aa}(t,t')_{\alpha,\alpha'} Q^{(1,1)}_{bb}(t',t)_{\alpha',\alpha} 
\nonumber \\
&\hspace{6em} 
- i \dfrac{g^2}{4} 
\left(
\bar{b}_\alpha^{(1)}(t) G_\psi^A(t,t') Q^{(2,1)}_{aa}(t,t')_{\alpha,\alpha'} b_{\alpha'}^{(1)}(t')
+
\bar{a}_\alpha^{(1)}(t) G_\psi^A(t',t) Q^{(1,1)}_{bb}(t,t')_{\alpha,\alpha'} a_{\alpha'}^{(2)}(t')\right)
\nonumber \\
&\hspace{4em} 
+ 
\dfrac{g^2}{8} G_\psi^A(t,t') Q^{(2,2)}_{aa}(t,t')_{\alpha,\alpha'} Q^{(2,1)}_{bb}(t',t)_{\alpha',\alpha} 
\nonumber \\
&\hspace{6em} 
- i \dfrac{g^2}{4} 
\left(
\bar{b}_\alpha^{(1)}(t) G_\psi^A(t,t') Q^{(2,2)}_{aa}(t,t')_{\alpha,\alpha'} b_{\alpha'}^{(2)}(t')
+
\bar{a}_\alpha^{(2)}(t) G_\psi^A(t',t) Q^{(2,1)}_{bb}(t,t')_{\alpha,\alpha'} a_{\alpha'}^{(2)}(t')
\right)
\nonumber \\
&\hspace{4em} 
+
\dfrac{g^2}{8} G_\psi^A(t,t') Q^{(1,1)}_{aa}(t,t')_{\alpha,\alpha'} Q^{(1,2)}_{bb}(t',t)_{\alpha',\alpha}
\nonumber \\
&\hspace{6em} 
- i \dfrac{g^2}{4} 
\left(
\bar{b}_\alpha^{(2)}(t) G_\psi^A(t,t') Q^{(1,1)}_{aa}(t,t')_{\alpha,\alpha'} b_{\alpha'}^{(1)}(t') 
+
\bar{a}_\alpha^{(1)}(t) G_\psi^A(t',t) Q^{(1,2)}_{bb}(t,t')_{\alpha,\alpha'} a_{\alpha'}^{(1)}(t')
\right) \nonumber \\
&\hspace{4em} 
+ 
\dfrac{g^2}{8} G_\psi^A(t,t') Q^{(1,2)}_{aa}(t,t')_{\alpha,\alpha'} Q^{(2,2)}_{bb}(t',t)_{\alpha',\alpha}
\nonumber \\
&\hspace{6em} 
- i \dfrac{g^2}{4} 
\left(
\bar{b}_\alpha^{(2)}(t) G_\psi^A(t,t') Q^{(1,2)}_{aa}(t,t')_{\alpha,\alpha'} b_{\alpha'}^{(2)}(t')
+
\bar{a}_\alpha^{(2)}(t) G_\psi^A(t',t) Q^{(2,2)}_{bb}(t,t')_{\alpha,\alpha'} a_{\alpha'}^{(1)}(t')
\right),
\end{align}}\twocolumngrid
Taking the terms linear in $Q$, we can rewrite this part of $S_{\phi\phi}$ in the style of (\ref{eq:Sphi}), such that
\begin{equation}
S_{\phi\phi}\bigg\vert_{\textrm{linear in $Q$}}
=
\iint \mathrm{d}t \mathrm{d}t' \sum_{\alpha,\alpha'} \bar{\phi}_\alpha(t) \Sigma(t,t') \phi_{\alpha'}(t').
\end{equation}
For non-equilibrium systems in the steady state, the two-time Green's functions are time invariant ($\tau=t-t'$) so self energy can be written in matrix form,
\begin{equation}
\boldsymbol{\Sigma} =
\begin{bmatrix}
\Sigma^{(1,1)}_{bb} & 0 & \Sigma^{(1,2)}_{bb} & 0 \\
0 & \Sigma^{(1,1)}_{aa} & 0 & \Sigma^{(1,2)}_{aa} \\
\Sigma^{(2,1)}_{bb} & 0 & \Sigma^{(2,2)}_{bb} & 0 \\
0 & \Sigma^{(2,1)}_{aa} & 0 & \Sigma^{(2,2)}_{aa} 
\end{bmatrix}
\end{equation}
with elements
\widetext{\allowdisplaybreaks\begin{align}
\Sigma^{(1,1)}_{bb}(\tau) &= -i\dfrac{g^2}{4}
\left(
G_\psi^K(\tau)Q^{(1,1)}_{aa}(\tau)_{\alpha,\alpha'} + 
G_\psi^R(\tau)Q^{(1,2)}_{aa}(\tau)_{\alpha,\alpha'} + 
G_\psi^A(\tau)Q^{(2,1)}_{aa}(\tau)_{\alpha,\alpha'}
\right)
\\
\Sigma^{(1,2)}_{bb}(\tau) &= -i\dfrac{g^2}{4}
\left(
G_\psi^K(\tau)Q^{(1,2)}_{aa}(\tau)_{\alpha,\alpha'} + 
G_\psi^R(\tau)Q^{(1,1)}_{aa}(\tau)_{\alpha,\alpha'} + 
G_\psi^A(\tau)Q^{(2,2)}_{aa}(\tau)_{\alpha,\alpha'}
\right)
\\
\Sigma^{(2,1)}_{bb}(\tau) &= -i\dfrac{g^2}{4}
\left(
G_\psi^K(\tau)Q^{(2,1)}_{aa}(\tau)_{\alpha,\alpha'} + 
G_\psi^R(\tau)Q^{(2,2)}_{aa}(\tau)_{\alpha,\alpha'} + 
G_\psi^A(\tau)Q^{(1,1)}_{aa}(\tau)_{\alpha,\alpha'}
\right)
\\
\Sigma^{(2,2)}_{bb}(\tau) &= -i\dfrac{g^2}{4}
\left(
G_\psi^K(\tau)Q^{(2,2)}_{aa}(\tau)_{\alpha,\alpha'} + 
G_\psi^R(\tau)Q^{(2,1)}_{aa}(\tau)_{\alpha,\alpha'} + 
G_\psi^A(\tau)Q^{(1,2)}_{aa}(\tau)_{\alpha,\alpha'}
\right)
\\
\Sigma^{(1,1)}_{aa}(\tau) &= -i\dfrac{g^2}{4}
\left(
G_\psi^K(-\tau)Q^{(1,1)}_{bb}(\tau)_{\alpha,\alpha'} + 
G_\psi^R(-\tau)Q^{(2,1)}_{bb}(\tau)_{\alpha,\alpha'} + 
G_\psi^A(-\tau)Q^{(1,2)}_{bb}(\tau)_{\alpha,\alpha'}
\right)
\\
\Sigma^{(1,2)}_{aa}(\tau) &= -i\dfrac{g^2}{4}
\left(
G_\psi^K(-\tau)Q^{(1,2)}_{bb}(\tau)_{\alpha,\alpha'} + 
G_\psi^R(-\tau)Q^{(2,2)}_{bb}(\tau)_{\alpha,\alpha'} + 
G_\psi^A(-\tau)Q^{(1,1)}_{bb}(\tau)_{\alpha,\alpha'}
\right)
\\
\Sigma^{(2,1)}_{aa}(\tau) &= -i\dfrac{g^2}{4}
\left(
G_\psi^K(-\tau)Q^{(2,1)}_{bb}(\tau)_{\alpha,\alpha'} + 
G_\psi^R(-\tau)Q^{(1,1)}_{bb}(\tau)_{\alpha,\alpha'} + 
G_\psi^A(-\tau)Q^{(2,2)}_{bb}(\tau)_{\alpha,\alpha'}
\right)
\\
\Sigma^{(2,2)}_{aa}(\tau) &= -i\dfrac{g^2}{4}
\left(
G_\psi^K(-\tau)Q^{(2,2)}_{bb}(\tau)_{\alpha,\alpha'} + 
G_\psi^R(-\tau)Q^{(1,2)}_{bb}(\tau)_{\alpha,\alpha'} + 
G_\psi^A(-\tau)Q^{(2,1)}_{bb}(\tau)_{\alpha,\alpha'}
\right),
\end{align}
}\twocolumngrid

This process shows that we make corrections to $\mathbf{G}^{-1}$ by producing an additional self-energy. By adding an interaction that we derive separately, we now construct an augmented action,
\widetext\begin{equation}
S_\textrm{aug} = \iint \mathrm{d}t \mathrm{d}t' \sum_{\alpha,\alpha'} \bar{\phi}_\alpha(t)
\mathbf{G}^{-1}_{\alpha,\alpha'}(t-t')
\phi_{\alpha'}(t') + S_Q + S_\psi
\end{equation}\twocolumngrid
Then integrating out the fermionic fields $\phi$ produces an effective action,
\begin{equation}
S_\textrm{aug}^\textrm{eff} = -i \sum_{\alpha} \mathrm{tr} \ln \mathbf{G}^{-1}_{\alpha,\alpha} + S_Q + S_\psi
\end{equation}
where tr traces over all parameters.

\subsection{Saddle-point analysis}

We calculate the mean field equation from the saddle points of the effective action \cite{Szymanska2007,Kamenev2011,Altland2010} taken relative to the classical and quantum fields, 
\begin{gather}
\dfrac{\delta S_\textrm{aug}^\textrm{eff}}{\delta \bar{\psi}_q} \bigg\vert_{\psi_q=0} = 0, 
\\
\dfrac{\delta S_\textrm{aug}^\textrm{eff}}{\delta \bar{\psi}_{cl}} \bigg\vert_{\psi_q=0} = 0 \quad \textrm{trivially}, 
\\
\dfrac{\delta S_\textrm{aug}^\textrm{eff}}{\delta Q} = 0.
\end{gather}
The $\delta S_\textrm{aug}^\textrm{eff}/\delta \bar{\psi}_q = 0$ saddle-point produces 
\begin{equation}
(\omega_0-\mu_S-i\kappa) \psi_f = i\dfrac{g}{2} \sum_\alpha \mathrm{tr} \; G^K_{ba}(\tau)_\alpha,
\end{equation}
where $\psi_f = \psi_{cl}/\sqrt{2}$. The functional derivative $\delta S_\textrm{aug}^\textrm{eff}/\delta Q = 0$ produces
\begin{equation}
\mathbf{Q}_{\alpha,\alpha'}(\tau) = -2 \mathbf{G}_{\alpha,\alpha'}(\tau).
\end{equation}
Using causality arguments of retarded and advanced Green's functions \cite{Kamenev2011} leads to
\begin{gather}
Q^{(1,1)} = Q^R, \\
Q^{(1,2)} = Q^K, \\
Q^{(2,1)} = 0, \\
Q^{(2,2)} = Q^A, \\
\Sigma^{(1,1)}_{bb} = \Sigma^R_{bb}, \\
\Sigma^{(1,2)}_{bb} = \Sigma^K_{bb}, \\
\Sigma^{(2,1)}_{bb} = \Sigma^{(2,1)}_{aa} = 0, \\
\Sigma^{(2,2)}_{bb} = \Sigma^A_{bb}, \\
\Sigma^{(1,1)}_{aa} = \Sigma^R_{aa}, \\
\Sigma^{(1,2)}_{aa} = \Sigma^K_{aa}, \\
\Sigma^{(2,2)}_{aa} = \Sigma^A_{aa}.
\end{gather}
We take the single frequency ansatz and get a self-consistent Dyson equation
\begin{equation}
\mathbf{G}(\omega) = \mathbf{G_0}(\omega) + \mathbf{G_0}(\omega) \cdot \boldsymbol{\Sigma}(\omega) \cdot \mathbf{G}(\omega).
\end{equation}
The self-energy 
\begin{gather}
\boldsymbol{\Sigma} =
\begin{bmatrix}
\Sigma^R_{bb} & 0 & \Sigma^K_{bb} & 0 \\
0 & \Sigma^R_{aa} & 0 & \Sigma^K_{aa} \\
0 & 0 & \Sigma^A_{bb} & 0 \\
0 & 0 & 0 & \Sigma^A_{aa} 
\end{bmatrix}
\end{gather}
in frequency is
\begin{align}
\Sigma^R_{bb}(\omega) &= 
i \dfrac{g^2}{4\pi}
\bigg( 
G^R_{aa}(\omega) \circ G_\psi^K(\omega) 
\nonumber\\
&\hspace{4em}
+ G^K_{aa}(\omega) \circ G_\psi^R(\omega) 
\bigg)
\\
\Sigma^R_{aa}(\omega) &= 
i \dfrac{g^2}{4\pi}
\bigg( 
G^R_{bb}(\omega) \circ G_\psi^K(-\omega) 
\nonumber\\
&\hspace{4em}
+ G^K_{bb}(\omega) \circ G_\psi^A(-\omega) 
\bigg)
\\
\Sigma^A_{bb}(\omega) &= 
i \dfrac{g^2}{4\pi}
\bigg(
G^A_{aa}(\omega) \circ G_\psi^K(\omega) 
\nonumber\\
&\hspace{4em}
+ G^K_{aa}(\omega) \circ G_\psi^A(\omega) 
\bigg)
\\
\Sigma^A_{aa}(\omega) &= 
i \dfrac{g^2}{4\pi}
\bigg(
G^A_{bb}(\omega) \circ G_\psi^K(-\omega) 
\nonumber\\
&\hspace{4em}
+ G^K_{bb}(\omega) \circ G_\psi^R(-\omega) 
\bigg)
\\
\Sigma^K_{bb}(\omega) &= 
i \dfrac{g^2}{4\pi}
\bigg(
G^K_{aa}(\omega) \circ G_\psi^K(\omega) 
\nonumber\\
&\hspace{4em}
+ G^R_{aa}(\omega) \circ G_\psi^R(\omega) 
\nonumber\\
&\hspace{4em}
+ G^A_{aa}(\omega) \circ G_\psi^A(\omega) 
\bigg)
\\
\Sigma^K_{aa}(\omega) &= 
i \dfrac{g^2}{4\pi}
\bigg(
G^K_{bb}(\omega) \circ G_\psi^K(-\omega) 
\nonumber\\
&\hspace{4em}
+ G^A_{bb}(\omega) \circ G_\psi^R(-\omega) 
\nonumber\\
&\hspace{4em}
+ G^R_{bb}(\omega) \circ G_\psi^A(-\omega)
\bigg)
\end{align}
where convolutions are defined as 
\begin{equation}
(f \circ g)(\omega) = \int \frac{\mathrm{d}\nu}{2\pi} f(\nu) g(\nu-\omega).
\end{equation}

\widetext

\begin{figure}[htpb]
\centering
\includegraphics[width=0.8\linewidth]{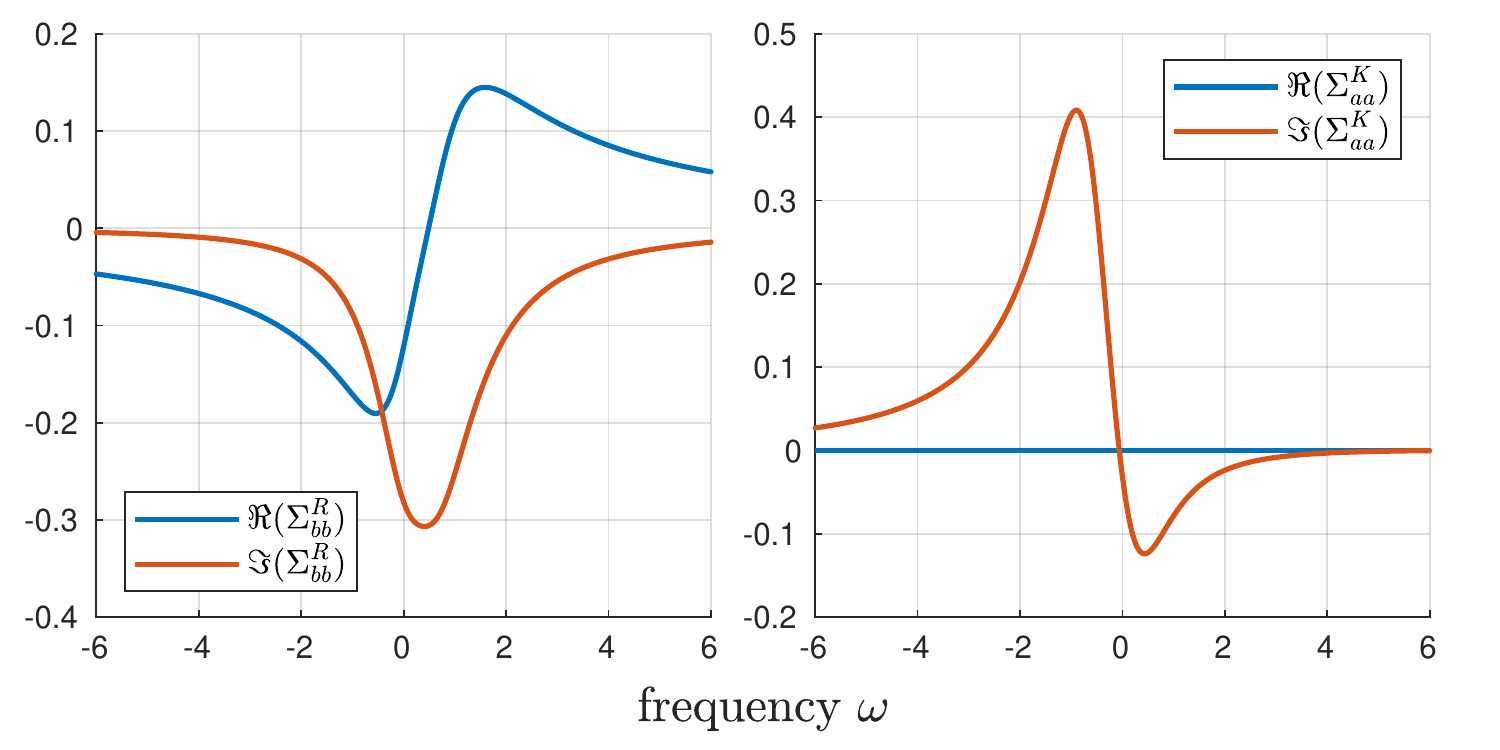}
\caption{Examples of self-energy $\Sigma$ as a function of frequency $\omega$.}
\label{fig:selfenergy}
\end{figure}

\twocolumngrid

As a first approximation, we let $\mathbf{G} = \mathbf{G_0}$. Then we numerically solve the saddle-point equation using Matlab's \texttt{fsolve} function. Through the trust region algorithm (either \texttt{trust-region} or \texttt{trust-region-dogleg}), we solve for the following nonlinear simultaneous equations
\begin{align}
(\omega_0-\mu_S)\psi_f - \Re 
\left(
i \dfrac{g}{2} \int \dfrac{\textrm{d}\omega}{2\pi} G^K_{ba}(\omega)_\alpha
\right) &= 0
\\
\kappa \psi_f - \Im
\left(
i \dfrac{g}{2} \int \dfrac{\textrm{d}\omega}{2\pi} G^K_{ba}(\omega)_\alpha
\right) &= 0 
\end{align}
which gives the numerical solution to $\mu_S$ and $\psi_f$. For simplicity, we set $g=1$, $\omega=2\epsilon_0=0$, $T_F=0.1g$.


\section{Normal state solutions}\label{app:normalstate} 

We now consider the stability of uniform condensed solutions. The consideration of stability is important because $\langle\psi\rangle=0$ is always a solution of the gap equation, so it is necessary to determine which of the normal and condensed solutions is stable. Secondly, because we considered only spatially homogeneous fields with a single oscillation frequency, it is possible that neither $\langle\psi\rangle=0$ nor the single-frequency ansatz is stable, suggesting more complex behaviour. 
There is an important difference in the interpretation of the saddle point equation between the closed-time-path path-integral formalism used here and the imaginary-time path integral in thermal equilibrium. In the imaginary time formalism, extremising the action corresponds to finding configurations which extremise the free energy. Thus, stable solutions correspond to a minimum of free energy, and unstable to local maxima. 

In contrast, for a classical saddle point (i.e., $\psi_q=0$), the action is always $S=0$ and the saddle point condition corresponds to configurations for which nearby paths add in phase. Thus, in order to study stability, one must directly investigate fluctuations about our ansatz, and determine whether such fluctuations grow or decay.
In considering the question of stability, we discuss stability of the normal state. This shows how the question of whether fluctuations about the non-equilibrium steady-state grow or decay is directly related to the instability expected in thermal equilibrium systems when the chemical potential goes above a bosonic mode. 
We consider an expansion in terms of $\psi=\langle\psi\rangle+\delta\psi$ to second order in $\delta\psi$. To find the spectrum of fluctuations, we consider the effective action governing fluctuations about either $\psi=\langle\psi\rangle$ or $\psi=0$. Considering the effective action
\begin{gather}
S_\textrm{aug}^\textrm{eff} = 
-i\sum_\alpha \mathrm{tr} (\ln \mathbf{G}^{-1}) + S_Q + S_\psi,
\\
\mathbf{G}^{-1} = \mathbf{G}_\mathbf{0}^{-1} - \boldsymbol{\Sigma}
\end{gather}
and expanding to second order in $\delta\psi$, one finds a contribution from the effective photon action, and a contribution from expanding the trace over fermions. Note $S_Q$ contains quadratic terms in $Q$, and $\boldsymbol{\Sigma}$ contains linear terms in $Q$. Using bosonic basis:
\begin{equation}
\mathbf{G}^{-1} = \mathbf{G}_\mathbf{0}^{-1} - \boldsymbol{\Sigma} =
\begin{bmatrix}
0 & (G^{-1}_0)^A \\ (G^{-1}_0)^R & (G^{-1}_0)^K
\end{bmatrix}
-
\begin{bmatrix}
0 & \Sigma^A \\ \Sigma^R & \Sigma^K
\end{bmatrix}.
\end{equation}
Expanding to second order in $\delta\psi$ gives contributions from $S_\psi$ and $-i\sum_\alpha \mathrm{tr} (\ln \mathbf{G}^{-1})$. The latter term gives
\begin{equation}
\mathbf{G}_{\alpha,\alpha}^{-1} = (\mathbf{G}^\mathbf{sp}_{\alpha,\alpha})^{-1} + \boldsymbol{\delta}\mathbf{G}_\alpha^{-1}, 
\end{equation}
where $\mathbf{G^\textrm{sp}}$ is the saddle-point Green's function, which depends on the value of the  saddle-point field $\langle\psi\rangle$. The contribution from fluctuations is
\widetext\begin{align}
\boldsymbol{\delta}\mathbf{G}_\alpha^{-1} 
&= \boldsymbol{\delta}\mathbf{G}^{-1}_{\textbf{0};\alpha} - \boldsymbol{\delta\Sigma}_{\alpha} \\
&=
\dfrac{g}{\sqrt{2}} 
\left(
(\delta\bar{\psi}_{q} \cdot \sigma_- + \delta\psi_{q} \cdot \sigma_+) \sigma_1^K
+ (\delta\bar{\psi}_{cl} \cdot \sigma_- + \delta\psi_{cl} \cdot \sigma_+) \sigma_0^K
\right)
-\boldsymbol{\delta\Sigma}_\alpha \\
&=
\dfrac{g}{\sqrt{2}} 
\left(
(\delta\bar{\psi}_{q} \cdot \sigma_- + \delta\psi_{q} \cdot \sigma_+) \sigma_1^K
+ (\delta\bar{\psi}_{cl} \cdot \sigma_- + \delta\psi_{cl} \cdot \sigma_+) \sigma_0^K
\right).
\end{align}\twocolumngrid
The full Dyson equation becomes
\begin{equation}
\mathbf{G}^\mathbf{sp}_{\alpha,\alpha} =
\mathbf{G}^\mathbf{sp}_{\mathbf{0};\alpha,\alpha} + \mathbf{G}^\mathbf{sp}_{\mathbf{0};\alpha,\alpha} \cdot \boldsymbol{\Sigma}_{\alpha,\alpha} \cdot \mathbf{G}^\mathbf{sp}_{\alpha,\alpha}
\end{equation}
and $\boldsymbol{\Sigma}_{\alpha,\alpha}$ contains convolutions of $\mathbf{G}^\mathbf{sp}_{\alpha,\alpha}$ and $\mathbf{G}_{\boldsymbol{\psi}}$. Hence the action can be rewritten by expanding
\widetext\begin{align}
-i \sum_\alpha \mathrm{tr}\ln(\mathbf{G}_{\alpha,\alpha}^{-1})
&= -i \sum_\alpha \mathrm{tr}\ln((\mathbf{G}^\mathbf{sp}_{\alpha,\alpha})^{-1}) 
- i \sum_\alpha \mathrm{tr}(\mathbf{G}^\mathbf{sp}_{\alpha,\alpha} \cdot \boldsymbol{\delta} \mathbf{G}_\alpha^{-1}) 
+ \dfrac{i}{2} \sum_\alpha \mathrm{tr}(\mathbf{G}^\mathbf{sp}_{\alpha,\alpha} \cdot \boldsymbol{\delta} \mathbf{G}_\alpha^{-1} \cdot \mathbf{G}^\mathbf{sp}_{\alpha,\alpha} \cdot \boldsymbol{\delta} \mathbf{G}_\alpha^{-1})
\\
&= -i \sum_\alpha \mathrm{tr}\ln((\mathbf{G}^\mathbf{sp}_{\alpha,\alpha})^{-1}) 
- i \sum_\alpha \mathrm{tr}(\mathbf{G}^\mathbf{sp}_{\alpha,\alpha} \cdot \boldsymbol{\delta} \mathbf{G}_{0;\alpha}^{-1}) 
+ \dfrac{i}{2} \sum_\alpha \mathrm{tr}(\mathbf{G}^\mathbf{sp}_{\alpha,\alpha} \cdot \boldsymbol{\delta} \mathbf{G}_{0;\alpha}^{-1} \cdot \mathbf{G}^\mathbf{sp}_{\alpha,\alpha} \cdot \boldsymbol{\delta} \mathbf{G}_{0;\alpha}^{-1})
\end{align}\twocolumngrid
and only retaining diagonal site index terms and neglecting bath-induced interaction terms between different $\alpha$ sites.

Now, we consider the quadratic terms in $\delta\psi$ with 
\begin{equation}
\delta\Lambda = 
\begin{bmatrix}
\delta\psi_{cl} (\omega) \\ \delta\bar{\psi}_{cl} (-\omega) \\
\delta\psi_{q} (\omega) \\ \delta\bar{\psi}_{q} (-\omega) 
\end{bmatrix}
\end{equation}
the action for fluctuations becomes
\begin{equation}
\delta S_f = \int \dfrac{\mathrm{d}\omega}{2\pi}
\delta\bar{\Lambda}(\omega)
\begin{bmatrix}
0 & (D^{-1})^A \\ (D^{-1})^R & (D^{-1})^K 
\end{bmatrix}
\delta\Lambda(\omega)
\end{equation}
where
\begin{equation}
(D^{-1})^{R,A,K} = 
\begin{bmatrix}
K_1^{R,A,K} & K_2^{R,A,K} \\ K_3^{R,A,K} & K_4^{R,A,K}
\end{bmatrix}.
\end{equation}

In the normal state, Nambu ($2 \times 2$) structure is redundant so the distribution function is a diagonal constant matrix $F_s=1+n_s$, where $n_s$ is the occupation of (bosonic) modes. There are also no anomalous (off-diagonal in Nambu space) contributions, and so  $K$ can be simplified using Keldysh Green's functions symmetry relations.

Taking the difference of luminescence and absorption gets the spectral weight
\widetext\begin{equation}
2\pi W(\omega,\mathbf{p}=0) =
\dfrac{\Im(K^R_1(\omega))}{\Re(K^R_1(\omega,\mathbf{p}=0))^2 + \Im(K^R_1(\omega))^2}
\end{equation}
where
\begin{align}
K^R_1(\omega) &=
\dfrac{1}{2}(\omega-\tilde{\omega}_0+i\kappa)
+ i \dfrac{g^2}{4} \int \dfrac{\mathrm{d}\omega'}{2\pi}
\left(
G^R_{bb}(\omega') G^K_{aa}(\omega'-\omega) + G^K_{bb}(\omega') G^A_{aa}(\omega'-\omega)
\right).
\end{align}\twocolumngrid

If the imaginary part of $K^R_1$ is a smooth function of $\omega$, then the distribution function is 
\begin{equation}
F_s(\omega) =
\dfrac{-i K^K_1(\omega)}{2\Im(K^R_1(\omega))}.
\end{equation}
If $\Im(K^R_1(\omega))=0$, then $F_s$ diverges. However the spectral weight $W=0$ so that number of photons does not diverge. Hence in order to find $\mu^\textrm{eff}$, it is necessary to solve for
\begin{equation}
\Im(K^R_1(\mu^\textrm{eff}))=0.
\end{equation}

\widetext

\begin{figure}[ht]
\centering
\subfloat[][Examples of $\Im(K^R_1(\omega))$ without incoherent drive.]{
\includegraphics[width=0.4\textwidth]{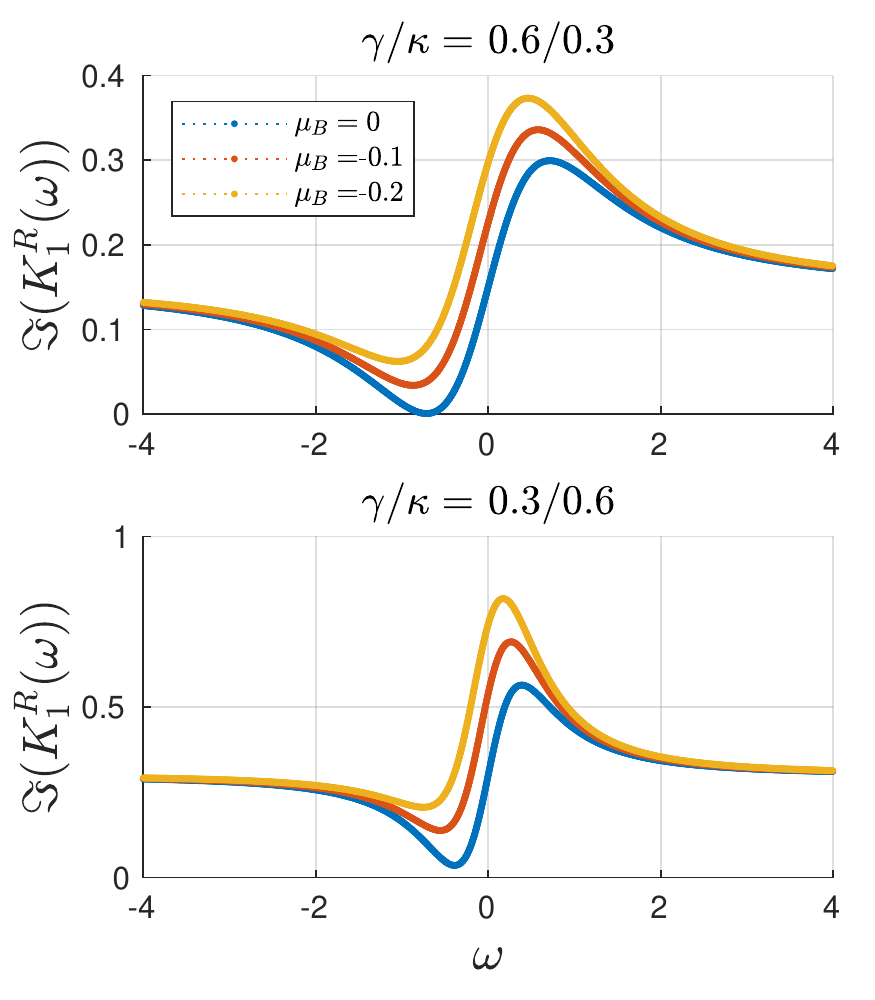}
\label{fig:kr1_paper}}
\subfloat[][Examples of $\Im(K^R_1(\omega))$ with incoherent drive.]{
\includegraphics[width=0.4\textwidth]{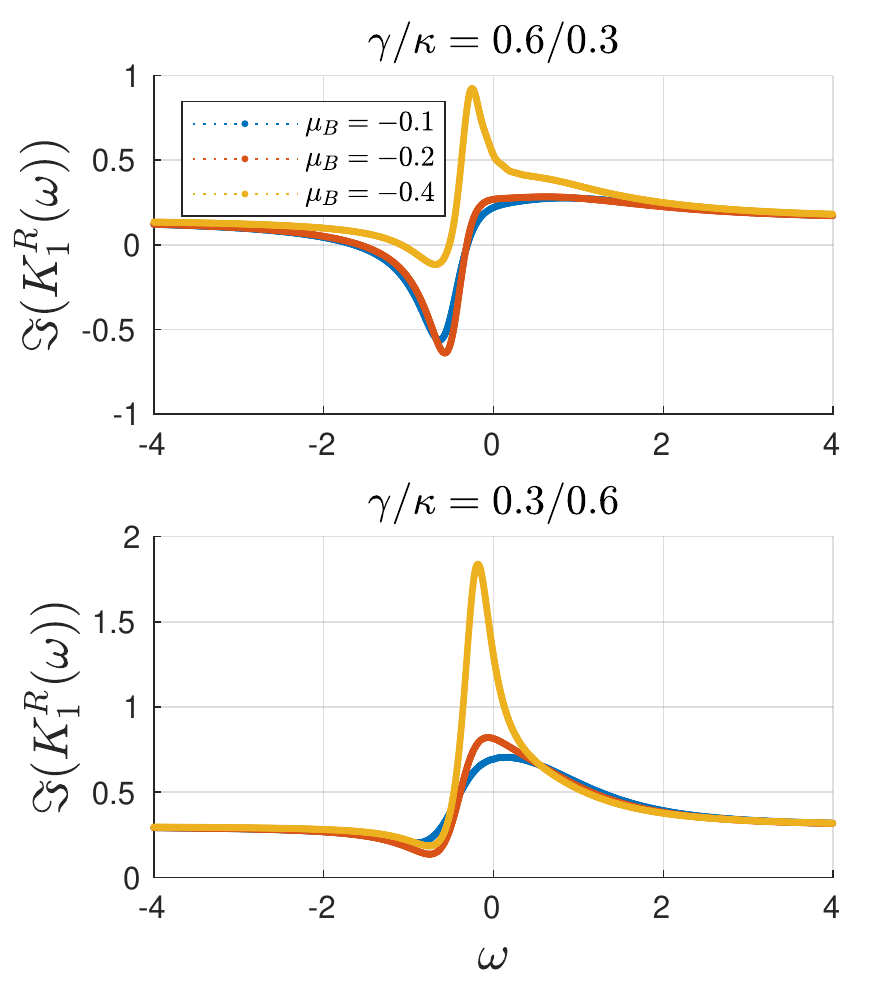}
\label{fig:kr1_full_paper}}
\caption{Normal state Green's function $\Im(K^R_1(\omega))$ as a function of frequency $\omega$ with and without incoherent drive. To determine $\mu^\textrm{eff}$, it is necessary to find where the graph crosses the horizontal axis.}
\end{figure}

\twocolumngrid

As a simple example, let $\mathbf{G}=\mathbf{G_0}$. When $\Im(K^R_1(\omega))$ is sketched as a function of frequency (Fig. \ref{fig:kr1_paper}), we find where the graph crosses the horizontal axis in order to determine $\mu^\textrm{eff}$. As the dephasing bath becomes stronger, the $\Im(K^R_1(\omega))$ graph is shifted upwards. This is observed as $\mu_B$ decreases to negative values: physically it means that increasingly more spins are in the ground state. At some critical $\mu_B$, the graph no longer crosses the horizontal axis, hence $\mu^\textrm{eff}$ does not exist. 
This problem also exists when we apply incoherent driving and define $K^R_1(\omega)$ in terms of $\mathbf{G} = \mathbf{G_0} + \mathbf{G_0} \cdot \boldsymbol{\Sigma} \cdot \mathbf{G}$ (Fig. \ref{fig:kr1_full_paper}). The graph shapes can change dramatically when the dephasing strength increases ($\mu_B$ decreases), compared to the simpler case without bosonic driving. This makes it more difficult to determine whether $\mu^\textrm{eff}$ exists, and if it is physical.

\end{document}